\documentstyle[12pt]{article}

\begin{document}

\begin{flushright}
IMSc/2013/12/08 
\end{flushright} 

\vspace{2mm}

\vspace{2ex}

\begin{center}



{\large \bf Stars in M theory} 

\vspace{4ex}

{\large \bf (made up of intersecting branes)} 

\vspace{8ex}

{\large  S. Kalyana Rama}

\vspace{3ex}

Institute of Mathematical Sciences, C. I. T. Campus, 

Tharamani, CHENNAI 600 113, India. 

\vspace{1ex}

email: krama@imsc.res.in \\ 

\end{center}

\vspace{6ex}

\centerline{ABSTRACT}

\begin{quote} 

We study stars in M theory. First, we obtain the analog of
Oppenheimer -- Volkoff equations in a suitably general set
up. We obtain analytically the asymptotic solutions to these
equations when the equations of state are linear. We study
perturbations around such solutions in several examples and,
following a standard method, use their behaviour to determine
whether an instability is present or not. In this way, we obtain
a generalisation of the corresponding results of Chavanis. We
also find that stars in M theory have instabilities. Therefore,
if sufficiently massive, such stars will collapse. We discuss
the significance of these (in)stabilities within the context of
Mathur's fuzz ball proposal.

\end{quote}

\vspace{2ex}


%
%

%
%
%
%
%
%
%

\newpage

\vspace{4ex}

\centerline{\bf 1. Introduction}  

\vspace{2ex}

Stars in our $3 + 1$ dimensional universe are expected to
collapse if they are sufficiently massive. Depending on their
masses, they may collapse to form compact objects like white dwarfs
or neutron stars, or collapse to form black holes. The compact
objects may collapse further if they gain sufficient mass, for
example, by accretion. Thus, ultimately, all sufficiently
massive objects are expected to collapse to form black holes.

Stars and other compact objects may be taken to be static and
spherically symmetric, with their constituents obeying
appropriate equations of state. The equilibrium configurations
are then determined by Oppenheimer -- Volkoff (OV) equations. By
studying perturbations around such equilibrium configurations,
one determines the onset of instability which signals the onset
of collapse \cite{c} -- \cite{shapiro}. Studying the collapse
further and determining the end state of collapse is
complicated, and is a field of ongoing research \cite{joshi}.
Nevertheless, a sufficiently massive object is generally assumed
to ultimately collapse to form a black hole.

It is clearly of interest to study stars and their collapse
dynamics in string/M theory. In general, for the same entropic
reasons which are explained in \cite{cm, k07} where early
universe was studied in string/M theory, we may assume that
stars in string/M theory are made up of ${\cal N}$ stacks of
intersecting branes. The equations of state for intersecting
branes given in these studies are applicable here also. Or, they
can be derived using similar methods. In the following, we will
restrict ourselves to M theory. The corresponding string theory
results are straightforward to obtain.

In this paper, we consider a $D = n_c + m + 2$ dimensional
spacetime where $n_c$ dimensional space is assumed to be compact
and toroidal, and the $(m + 1)$ dimensional transverse space,
with $m \ge 2$, is assumed to be non compact. We assume suitable
isometries along the compact space and, further, that the stars
are made up of ${\cal N}$ number of non interacting
multicomponent fluids. This set up, for suitable choices of $(D,
n_c, {\cal N})$ and suitable equations of state, is applicable
to the stars made up of intersecting branes in M theory.

We then obtain the analog of OV equations in this general set
up. Solving these OV type equations generally and analytically
is not possible. The standard procedure is to resort to
numerical methods.

However, following the methods of \cite{cpaper, p, pc02, pc07},
it is possible to determine analytically whether an instability
is present or not. In this method, one first obtains the
asymptotic solutions of the OV equations. Such solutions,
referred to in \cite{cpaper} as singular solutions, can be
obtained analytically when the equations of state are
linear. One then studies the behaviour of the perturbations in
the mass of the star around these asymptotic solutions. This
behaviour is sufficient to determine the presence or absence of
the instability: if the perturbations exhibit damped
oscillations then an instability is present; and if they exhibit
a monotonous behaviour with no oscillations then no instability
is present. Technically, determining which behaviour is
exhibited by the perturbations involves determining the sign of
a certain discriminant. See \cite{p, lw, y, katz} for detailed
explanations of a similar instability, called gravothermal
catastrophe, which occur in isothermal spheres.

We follow this method. Taking the equations of state to be
linear, we obtain analytically the asymptotic solutions to the
OV type equations. We then study the perturbations around the
asymptotic solutions. A general answer regarding the presence or
absence of an instabilty is algebraically quite involved. Hence,
and in order to illustrate the nature of the results, we
consider a few select examples only and study the perturbations.
These examples are: (1) Stars which are effectively same as
those in \cite{pc07}. (2) Stars for which ${\cal N} = 1$ and
$n_c$ is arbitrary. (3) Stars made up of $M2$ branes, or $M5$
branes. And, (4) stars made up of four stacks of intersecting
branes, denoted as $2 2' 5 5'$ configuration. Similar results
hold for stars made up of three stacks of $2 2' 2''$
intersecting brane configuration.

Our results are a generalisation of those of \cite{pc07} and are
qualitatively similar. Thus, {\em e.g.} in Example (2) with the
equation of state being given by $p = w \rho$ where $0 \le w \le
1 \;$, we find that there is no instability if $m \ge 9
\;$. Thus in these dimensions, there will not be any collpase,
and a stable equilibrium configuration is possible for any value
of mass. In the other examples of stars in M theory, $n_c$ and
$m$ are fixed and $D = 11$, and instabilities are present. Thus,
a sufficiently massive stars in these examples will collapse.

We then discuss the significance of these (in)stability results,
particularly within the context of Mathur's fuzzball proposal
\cite{fuzz}.

This paper is organised as follows. In section {\bf 2}, we
present the general set up and write down the analog of OV
equations. In section {\bf 3}, we incorporate the linear
equations of state. In section {\bf 4}, we obtain the asymptotic
solutions and study the perturbations.  In section {\bf 5}, we
specialise to the stars in M theory, obtain asymptotic
solutions, and study the perturbations. In section {\bf 6}, we
discuss the significance of the (in)stability results. In
section {\bf 7}, we conclude with a brief summary and mention a
few topics for further studies. In three appendices, we provide
some useful formulas.


\vspace{4ex}

\centerline{\bf 2. General Set Up}  

\vspace{2ex}

We are interested in studying the static solutions describing
stars made up of ${\cal N}$ stacks of M theory branes,
intersecting according to the BPS rules whereby two stacks of
five branes intersect along three common spatial directions; a
stack each of two branes and five branes intersect along one
common spatial direction; and two stacks of two branes intersect
along zero common spatial direction. We model these intersecting
M theory branes by ${\cal N}$ seperately conserved energy
momentum tensors $T_{M N (I)}$ and appropriate equations of
state among their components. We write the equations of motion
and an appropriate ansatz for the metric which can describe such
static configurations of M theory stars.

We take the spatial directions of the brane worldvolumes to be
toroidal and assume necessary isometries. Let the spacetime
coordinates be given by $x^M = (t, x^i, r, \theta^a)$ where $i =
1, 2, \cdots, n_c$ and $a = 1, 2, \cdots, m \;$. The total
spacetime dimension $D = n_c + m + 2$ which $= 11$ for stars in
M theory. The coordinates $x^i$ describe the compact, $n_c$
dimensional, toroidal space; and the radial coordinate $r$ and
the spherical coordinates $\theta^a$ describe the non compact,
$(m + 1)$ dimensional, transverse space. In the following, we
will assume that $m \ge 2 \;$. In standard notation and with
$\kappa^2 = 8 \pi G_D = 1 \;$, the equations of motion may be
written as
\begin{eqnarray}
{\cal R}_{M N} - \frac{1}{2} \; g_{M N}\; {\cal R} & = &
T_{M N} \; = \; \sum_I T_{M N (I)} \label{rmn} \\
\sum_M \; \nabla_M \; T^M_{\; \; N (I)} & = & 0 \label{tmnI}
\end{eqnarray}
where $T_{M N}$ is the total energy momentum tensor of non
interacting multicomponent fluids and $T_{M N (I)}$ the energy
momentum tensor for the $I^{th}$ component fluid, $I = 1, 2,
\cdots, {\cal N} \;$. For stars in M theory, $T_{M N}$ is the
total energy momentum tensor for intersecting branes and $T_{M N
(I)}$ is the energy momentum tensor for the $I^{th}$ stack of
branes, $I = 1, 2, \cdots, {\cal N} \;$.


In the following, we study the static solutions which are
spherically symmetric in the $(m + 1)$ dimensional transverse
space. The line element $d s$ which can describe such static
intersecting branes may now be written as
\begin{equation}\label{ds} 
d s^2 = g_{M N} \; d x^M \; d x^N 
= - e^{2 \lambda^0} d t^2 + \sum_i e^{2 \lambda^i} (d x^i)^2
+ e^{2 \lambda} d r^2 + e^{2 \sigma} d \Omega_m^2
\end{equation}
where $d \Omega_m$ is the standard line element on an $m$
dimensional unit sphere. 
The energy momentum tensors $T_{M N (I)}$ may all be assumed to
be diagonal. We denote these diagonal elements as

\begin{equation}\label{rhopiI}
\left( T^0_{\; \; 0 (I)}, \; T^i_{\; \; i (I)}, \; 
T^r_{\; \; r (I)}, \; T^a_{\; \; a (I)} \right) = 
\left( p_{0 I}, \; p_{i I}, \; \Pi_I, \; p_{a I} \right) 
\end{equation}
where $p_{0 I} = - \rho_I$ and $p_{a I} = p_{\Omega I}$ for all
$a \;$. The total energy momentum tensor is now given by
$T^M_{\; \; \; \; N} = diag \; \left( p_0, \; p_i, \; \Pi, \;
p_a \right)$ where $p_0 = \sum_I p_{0 I} = - \rho$ and
\begin{equation}\label{rhopitotal}
\rho = \sum_I \rho_I \; \; , \; \; \; 
p_i = \sum_I p_{i I} \; \; , \; \; \; 
\Pi = \sum_I \Pi_I \; \; , \; \; \; 
p_a = p_\Omega = \sum_I p_{\Omega I} \; \; .
\end{equation}
In the case of the stars being studied here, one also has $\Pi_I
= p_{a I} = p_{\Omega I} \;$.


\vspace{2ex}

\centerline{\bf Equations of motion}

\vspace{2ex}

Let
\begin{equation}\label{palphas} 
\alpha = (0, i, a) 
\; \; , \; \; \; 
\lambda^\alpha = (\lambda^0, \lambda^i, \lambda^a) 
\; \; , \; \; \; 
p_{\alpha I} = (p_{0 I}, p_{i I}, p_{a I}) 
\end{equation}
where $p_{0 I} = - \rho_I$ and
\[
\lambda^a = \sigma \; \; , \; \; \; 
p_{a I} = p_{\Omega I} = p_I  
\]
for all $a \;$. For the static solutions which are spherically
symmetric in the $(m + 1)$ dimensional transverse space, the
fields $(\lambda^\alpha, \; \lambda)$ and $(p_{\alpha I},
\Pi_I)$ are all assumed to depend only on the coordinate $r
\;$. Define
\begin{eqnarray}
\Lambda & = & \sum_\alpha \lambda^\alpha = 
\lambda^0 + \sum_i \lambda^i + m \sigma \label{Lambda} \\
T_I & = & \sum_M T^M_{\; \; \; M (I)} = 
\Pi_I + \sum_\alpha p_{\alpha I} \; \; . \label{T}
\end{eqnarray}
It is straightforward to calculate the Riemann tensor components
corresponding to the metric given by equation (\ref{ds}), and
obtain the equations of motion. Using the above definitions and
after some algebra, it follows that
\begin{eqnarray}
(\Pi_I)_r & = & - \Pi_I \; \Lambda_r 
+ \sum_\alpha p_{\alpha I} \lambda^\alpha_r \label{Pir} \\
\Lambda^2_r - \sum_\alpha (\lambda^\alpha_r)^2 & = &
2 \sum_I \Pi_I \; e^{2 \lambda} 
+ m (m - 1) \; e^{2 \lambda - 2 \sigma} \label{Lambdar2}
\end{eqnarray}
\begin{equation}\label{alpharr}
\lambda^\alpha_{r r} + (\Lambda_r - \lambda_r) \;
\lambda^\alpha_r = \sum_I \left(- p_{\alpha I} 
+ \frac{T_I}{D - 2} \right) \; e^{2 \lambda} 
+ \delta^{\alpha a} \; (m - 1) \; e^{2 \lambda - 2 \sigma}
\end{equation}
where the subscripts $r$ denote $r-$derivatives. We define a
function $f(r)$ by
\begin{equation}\label{f}
e^{2 \lambda - 2 \sigma} = \frac{1}{r^2 f} 
\end{equation}
and a mass function $M(r)$ by
\begin{equation}\label{M(r)}
M(r) = r^{m - 1} \; (1 - f) \;\; \; \longleftrightarrow \; \; \;
f = 1 - \frac{M}{r^{m - 1}} \; \; .
\end{equation}
Note that either of the functions $f$ and $M$ may be traded for
the function $\lambda (r) \;$, and that the line element for the
$(m + 1)$ dimensional transverse space given in equation
(\ref{ds}), written in terms of $f$, becomes
\begin{equation}\label{fdsm+1} 
e^{2 \lambda} d r^2 + e^{2 \sigma} d \Omega_m^2 \; = \; 
e^{2 \sigma} \; \left( \frac{d r^2}{r^2 f} + d \Omega_m^2
\right) \; \; . 
\end{equation}
Also, by a suitable change of varible, the equations of motion
(\ref{Pir}) -- (\ref{alpharr}) can be written more compactly as
shown in Appendix {\bf A}.


\vspace{2ex}

\centerline{\bf Reduction to $d = m + 2$ dimensions}

\vspace{2ex}

We will now dimensionally reduce on the compact $n_c$
dimensional toroidal space from $D = n_c + m + 2$ dimensions to
$d = m + 2$ dimensions. Let $x^M = (t, x^i, r, \theta^a)$ be the
$D$ dimensional spacetime coordinates as before and $x^\mu = (t,
r, \theta^a)$ be the $d$ dimensional spacetime coordinates.
Consider the $D$ dimensional line element $d s$ given by
equation (\ref{ds}), and denote its $d$ dimensional part as 
follows:
\begin{equation}\label{ds1d}
d s^2_d = g_{\mu \nu (d)} \; d x^\mu \; d x^\nu = 
- e^{2 \lambda^0} d t^2 + e^{2 \lambda} d r^2 
+ e^{2 \sigma} d \Omega_m^2 \; \; . 
\end{equation}

Upon dimensional reduction, symbolically, we have the following:
\begin{eqnarray*}
S & \sim & \int d^D x \; \sqrt{- g} \; R  \\
& \sim & \int d^d x \; \sqrt{- g_{(d)}} \; e^{\Lambda^c} \;
(R_{(d)} + \cdots) 
\; \; \; , \; \; \; \; \Lambda^c = \sum_i \lambda^i \\
& \sim & \int d^d x \; \sqrt{- \tilde{g}} \; 
(\tilde{R} + \cdots) 
\; \; \; , \; \; \; \; \; \; \; \tilde{g}_{\mu \nu} =
e^{\frac{2 \Lambda^c}{m}} \; g_{\mu \nu (d)}
\end{eqnarray*}
where $\tilde{g}_{\mu \nu}$ is the $d$ dimensional Einstein
frame metric. Hence the line element $\tilde{d s}_d$ in the $d$
dimensional Einstein frame is given by
\begin{equation}\label{dsd}
\tilde{d s}^2_d = \tilde{g}_{\mu \nu} \; d x^\mu \; d x^\nu = 
- e^{2 \tilde{\lambda}^0} d t^2 + e^{2 \tilde{\lambda}} d r^2 
+ e^{2 \tilde{\sigma}} d \Omega_m^2 
\end{equation}
where 
\begin{equation}\label{ltil1}
\tilde{\lambda}^\alpha = \lambda^\alpha + \frac{\Lambda^c}{m}
\; \; \; , \; \; \; \; 
\tilde{\lambda} = \lambda + \frac{\Lambda^c}{m} \; \; , 
\end{equation}
and $\tilde{\lambda}^a = \tilde{\sigma}$ for all $a \;$. Also,
let
\begin{equation}
\tilde{\Lambda} = \tilde{\lambda}^0 + m \tilde{\sigma} 
\; \; , \; \; \; 
\tilde{\chi} = \tilde{\Lambda} - \tilde{\sigma}
= \tilde{\lambda}^0 + (m - 1) \tilde{\sigma} \; \; . 
\end{equation}
We then have, with the functions $\chi$ and $\tau$ as defined in
Appendix {\bf A}, 
\[
\tilde{\Lambda} = \Lambda + \frac{\Lambda^c}{m} \; \; , \; \; \;
\tilde{\chi} = \chi = \Lambda - \sigma 
\]
and 
\[
e^{2 \lambda - 2 \sigma} = 
e^{2 \tilde{\lambda} - 2 \tilde{\sigma}} = \frac{1}{r^2 f}
\; \; \; , \; \; \; \; 
r_\tau = \frac{d r}{d \tau} = e^{\Lambda - \lambda} = 
e^{\tilde{\Lambda} - \tilde{\lambda}} = r \; \sqrt{f} \; e^\chi
\; \; .
\]
Note that it is equally convenient to use $\tilde{\lambda}^i$ or
$\lambda^i \;$; that $\Lambda^c$ is related to
$\tilde{\lambda}^i$ as follows: $\tilde{\Lambda}^c = \sum_i
\tilde{\lambda}^i = \frac{n_c + m} {m} \; \Lambda^c \;$; that
$\sum_i \tilde{a}^i b^i = \sum_i a^i \tilde{b}^i = \sum_i a^i
b^i + \frac{a^c b^c}{m}$ where $\tilde{a}^i = a^i + \frac{a^c}
{m}$, $\; a^c = \sum_i a^i$, and similarly for $\tilde{b}^i$ and
$b^c \;$; and, lastly, that $\sum_i \tilde{a}^i a^i > 0$ if
$a^i$ do not all vanish.

\vspace{1ex}

It now follows from equations (\ref{Pir}) -- (\ref{alpharr})
that
\begin{equation}
(\Pi_I)_r = - \Pi_I \; \tilde{\Lambda}_r 
+ p_{0 I} \; \tilde{\lambda}^0_r 
+ m \; p_{\Omega I} \; \tilde{\sigma}_r 
+ \sum_i \left( p_{i I} - \frac{{\cal T}_I}{m} \right)
\lambda^i_r 
+ \Pi_I \left( \frac{2 \Lambda^c_r}{m} \right)
\label{tldPir}
\end{equation}
\begin{eqnarray}
2 \tilde{\lambda}^0_r \tilde{\sigma}_r 
+ (m - 1) (\tilde{\sigma}_r)^2 & = &
\frac{2}{m} \sum_I \Pi_I \; e^{2 \lambda} 
+ (m - 1) \; e^{2 \tilde{\lambda} - 2 \tilde{\sigma}} 
+ \frac{{\cal B}}{m} \label{tldLambdar2} \\
\tilde{\sigma}_{r r} + (\tilde{\Lambda}_r - \tilde{\lambda}_r)
\; \tilde{\sigma}_r & = & \sum_I \left(- p_{\Omega I} 
+ \frac{{\cal T}_I}{m} \right) \; e^{2 \lambda} + (m - 1) \;
e^{2 \tilde{\lambda} - 2 \tilde{\sigma}} \label{tldsigmarr} \\
\tilde{\lambda}^0_{r r} + (\tilde{\Lambda}_r 
- \tilde{\lambda}_r) \; \tilde{\lambda}^0_r & = & 
\sum_I \left(- p_{0 I} + \frac{{\cal T}_I}{m} \right) \; 
e^{2 \lambda} \label{tld0rr} \\
\tilde{\lambda}^i_{r r} + (\tilde{\Lambda}_r 
- \tilde{\lambda}_r) \; \tilde{\lambda}^i_r & = & 
\sum_I \left(- p_{i I} + \frac{{\cal T}_I}{m} \right) \; 
e^{2 \lambda} \label{tldirr} 
\end{eqnarray}
where ${\cal T}_I = \Pi_I + p_{0 I} + m \; p_{\Omega I}$ and
${\cal B} = \sum_i (\lambda^i_r)^2 + \frac{(\Lambda^c_r)^2}{m} =
\sum_i \tilde{\lambda}^i_r \; \lambda^i_r \; $. Using the
diffeomorphic freedom in defining the radial coordinate, we now
set $e^{\tilde{\sigma}} = r \;$. Equations (\ref{tldLambdar2})
and (\ref{tldsigmarr}) become
\begin{eqnarray}
r \; \tilde{\lambda}^0_r & = & \sum_I \frac{\Pi_I}{m} \; r^2
\; e^{2 \lambda} + \frac{m - 1}{2} \; ( 
e^{2 \tilde{\lambda}} - 1 ) + \frac{r^2 {\cal B}}{2 m} 
\label{2tldLambdar2} \\
& & \nonumber \\
r \; (\tilde{\lambda}^0_r - \tilde{\lambda}_r) & = & \sum_I
\left(\frac{\Pi_I + p_{0 I}}{m} \right) \; r^2 \; e^{2 \lambda}
+ (m - 1) \; (e^{2 \tilde{\lambda}} - 1 ) \; \; . 
\label{2tldsigmarr} 
\end{eqnarray}


\vspace{2ex}

\centerline{\bf Stars}

\vspace{2ex}

For the stars, we have $\Pi_I = p_{a I} = p_I \;$. Then ${\cal
T}_I = ( m + 1) p_I - \rho_I \;$, and equations (\ref{tldPir})
and (\ref{tld0rr}) -- (\ref{2tldsigmarr}) become
\begin{equation}\label{*pir}
(p_I)_r = - (\rho_I + p_I) \; \tilde{\lambda}^0_r + \sum_i
\left( p_{i I} - \frac{( m + 1) p_I - \rho_I} {m} \right) \;
\lambda^i_r + p_I \left( \frac{2 \Lambda^c_r}{m} \right)
\end{equation}
\begin{eqnarray}
\tilde{\lambda}^0_{r r} + (\tilde{\Lambda}_r 
- \tilde{\lambda}_r) \; \tilde{\lambda}^0_r & = & 
\sum_I \left(\rho_I + \frac{( m + 1) p_I - \rho_I} {m} \right)
\; e^{2 \lambda} \label{*0rr} \\
\tilde{\lambda}^i_{r r} + (\tilde{\Lambda}_r 
- \tilde{\lambda}_r) \; \tilde{\lambda}^i_r & = & 
\sum_I \left(- p_{i I} + \frac{( m + 1) p_I - \rho_I} {m}
\right) \; e^{2 \lambda} \label{*irr}
\end{eqnarray}
\begin{equation}\label{*r2} 
r \; \tilde{\lambda}^0_r \; = \; \sum_I \frac{p_I}{m} \; r^2
\; e^{2 \lambda} + \frac{m - 1}{2} \; ( 
e^{2 \tilde{\lambda}} - 1 ) + \frac{r^2 {\cal B}}{2 m} 
\end{equation}
\begin{equation}\label{*sigmarr} 
r \; (\tilde{\lambda}^0_r - \tilde{\lambda}_r) \; = \; \sum_I
\left(\frac{p_I - \rho_I}{m} \right) \; r^2 \; e^{2 \lambda}
+ (m - 1) \; (e^{2 \tilde{\lambda}} - 1 ) \; \; .
\end{equation}
Since $e^{\tilde{\sigma}} = r \;$, we also have
\[
\tilde{d s}^2_d = - e^{2 \tilde{\lambda}^0} d t^2 
+ \frac{d r^2}{f} + r^2 d \Omega_m^2
\; \; \; , \; \; \; 
f = e^{- 2 \tilde{\lambda}} = 1 - \frac{M}{r^{m - 1}} \; \; .
\] 
Thus, the line element $\tilde{d s}_d$ in the $d = m + 2$
dimensional Einstein frame takes the standard form. Therefore
the functions appearing in it, {\em e.g.} the mass function $M$,
may be interpreted in the standard way. Note that equations
(\ref{2tldLambdar2}) and (\ref{2tldsigmarr}), equivalently
equations (\ref{*r2}) and (\ref{*sigmarr}), give
\[
r \; \tilde{\lambda}_r \; = \; \sum_I \frac{\rho_I}{m} \; r^2 \;
e^{2 \lambda} - \frac{m - 1}{2} \; ( e^{2 \tilde{\lambda}}
- 1 ) + \frac{r^2 {\cal B}}{2 m} \; \; .
\]
The definition $M(r) = r^{m - 1} \; ( 1 - f ) = r^{m - 1} \; ( 1
- e^{- 2 \tilde{\lambda}} ) \;$ then gives
\begin{equation}\label{mr}
M_r \; = \; \frac{r^m}{m} \; \left( 2 \sum_I \rho_I \; 
e^{- \frac{2 \Lambda^c}{m}} + {\cal B} \; 
e^{- 2 \tilde{\lambda}} \right) 
\end{equation} 
which leads to the familiar expression $M = \frac{2}{m} \;
\int^r_0 d r \; (r^m \rho) $ when $\lambda^i = 0 \;$.

Equations (\ref{*pir}) -- (\ref{mr}) are the analog of OV
equations. They describe the static equilibrium configurations
of stars in $D = n_c + m + 2$ dimensional spacetime. These
configurations are independent of the $n_c$ dimensional compact
toroidal coordinates and are spherically symmetric in the $(m +
1)$ dimensional transverse space. The corresponding equations
when $n_c = 0$, and hence $D = m + 2 \;$, are given in
\cite{pc07} and they follow from the above ones by formally
setting $\lambda^i = 0$, taking ${\cal N} = 1$, and ignoring the
$\tilde{\lambda}^i_{r r}$ equation. The standard four
dimensional OV equations follow upon further setting $m = 2 \;$.


\vspace{4ex}

\centerline{\bf 3. Linear equations of state}  

\vspace{2ex}

To solve equations (\ref{*pir}) -- (\ref{*sigmarr}) and obtain
the solutions for the fields $(\lambda^\alpha, \; p_{\alpha I},
\; \Pi_I)$, one further requires equations of state which give
$p_{\alpha I}$ and $\Pi_I$ as functions of $\rho_I \;$. In this
paper, we will take the equations of state to be linear and
write them as
\begin{equation}\label{eoswi}
p_{\alpha I} = w^I_\alpha \; \rho_I \; \; , \; \; \; 
\Pi_I = w^I_\Pi \; \rho_I
\end{equation}
where $(w^I_\alpha, \; w^I_\Pi)$ are constants, $w^I_0 = - 1$
since $p_{0 I} = - \rho_I$, and $w^I_a = w^I_\Omega = w^I$ since
$p_{a I} = p_{\Omega I} = p_I$ for all $a \;$. We show in
Appendix {\bf B} that when the equations of state are linear,
the fields $(\lambda^\alpha, \; \rho_I)$ can be expressed in
terms of ${\cal N} + 1$ independent fields, denoted as $(l^I, \;
l^*) \;$. One then has to solve the equations for $l^I$ and
$l^*$ only.


For the stars we study here, $w^I_\Pi = w^I_a = w^I$ since
$\Pi_I = p_{a I} = p_I \;$. Define
\begin{eqnarray}
c^{\alpha I} & = & - w^I_\alpha + \frac{(m + 1) w^I 
+ \sum_j w^I_j - 1}{n_c + m} \label{calpha} \\
& & \nonumber \\
\tilde{c}^{\alpha I} & = & c^{\alpha I} 
+ \frac{ \sum_j c^{j I}}{m}
\; = \; - w^I_\alpha + \frac{(m + 1) w^I - 1}{m} \; \; .
\label{tldcalpha} 
\end{eqnarray}
Thus $\tilde{c}^{a I} = \tilde{c}^I$ and
\[
\tilde{c}^{0 I} = 1 + w^I + \tilde{c}^I 
\; \; , \; \; \;
\tilde{c}^{i I} = - w^I_i + w^I + \tilde{c}^I
\; \; , \; \; \;
\tilde{c}^I = \frac{w^I - 1}{m} \; \; . 
\]
Now consider equation (\ref{*pir}). Upon using $p_{\alpha I} =
w^I_\alpha \; \rho_I \;$, it gives
\begin{equation}\label{wrhoir}
w^I \; (ln \; \rho_I)_r \; = \; - \; (1 + w^I) \;
\tilde{\lambda}^0_r \; - \; \sum_i \tilde{c}^{i I} \;
\lambda^i_r + \; w^I \; \left( \frac{2 \Lambda^c_r}{m} \right)
\end{equation}
from which it follows that 
\begin{equation}\label{phii}
\rho_I = \rho_{I 0} \; e^{\phi^I} \; e^{\frac{2 \Lambda^c}{m}} 
\; \; \; , \; \; \; \; 
w^I \; \phi^I 
\; = \; - \; (1 + w^I) \; \tilde{\lambda}^0
- \sum_i \tilde{c}^{i I} \; \lambda^i 
\end{equation}
where $\rho_{I 0}$ is a constant. Therefore
\[
\rho_I \; e^{2 \lambda} = \rho_{I 0} \; e^{\phi^I 
+ 2 \tilde{\lambda}}
\; \; \; , \; \; \; \; 
r^2 \rho_I \; e^{2 \lambda} = \rho_{I 0} \; e^{\phi^I 
+ 2 \tilde{\lambda} + 2 \tilde{\sigma}} \; \; .
\]

For any function $X(r(\tilde{\sigma}))$, using $r =
e^{\tilde{\sigma}}$, we have
\[ 
X_{\tilde{\sigma}} = r X_r \; \; , \; \; \; 
X_{\tilde{\sigma} \tilde{\sigma}} = r^2 X_{r r} + r X_r
\]
where the subscripts $\tilde{\sigma}$ denote
$\tilde{\sigma}-$derivatives. Then
\[
r^2 \left( X_{r r} + (\tilde{\Lambda}_r - \tilde{\lambda}_r)
\; X_r \right) \; = \; 
X_{\tilde{\sigma} \tilde{\sigma}} + 
(\tilde{\chi}_{\tilde{\sigma}} -
\tilde{\lambda}_{\tilde{\sigma}}) \; X_{\tilde{\sigma}}
\]
where $ \tilde{\chi} = \tilde{\Lambda} - \tilde{\sigma} =
\tilde{\lambda}^0 + (m - 1) \; \tilde{\sigma} \;$. Equations
(\ref{*0rr}) -- (\ref{*sigmarr}), written in terms of
$\tilde{\sigma}$, now become
\begin{eqnarray}
\tilde{\lambda}^0_{\tilde{\sigma} \tilde{\sigma}} 
+ (\tilde{\chi}_{\tilde{\sigma}} 
- \tilde{\lambda}_{\tilde{\sigma}}) \; 
\tilde{\lambda}^0_{\tilde{\sigma}} & = & 
\sum_I \tilde{c}^{0 I} \; \rho_{I 0} \; e^{\phi^I 
+ 2 \tilde{\lambda} + 2 \tilde{\sigma}} \label{w0rr} \\
\tilde{\lambda}^i_{\tilde{\sigma} \tilde{\sigma}} 
+ (\tilde{\chi}_{\tilde{\sigma}} 
- \tilde{\lambda}_{\tilde{\sigma}}) \; 
\tilde{\lambda}^i_{\tilde{\sigma}} & = & 
\sum_I \tilde{c}^{i I} \; 
\rho_{I 0} \; e^{\phi^I + 2 \tilde{\lambda} 
+ 2 \tilde{\sigma}} \label{wirr} 
\end{eqnarray}
\begin{equation}\label{wquad} 
\tilde{\lambda}^0_{\tilde{\sigma}} \; = \; \sum_I \frac{w^I}{m}
\; \rho_{I 0} \; e^{\phi^I + 2 \tilde{\lambda} 
+ 2 \tilde{\sigma}} + \frac{m - 1}{2} \; \left( 
e^{2 \tilde{\lambda}} - 1 \right) + \frac{r^2 {\cal B}}{2 m}
\end{equation}
\begin{equation}\label{wsigma} 
(\tilde{\lambda}^0_{\tilde{\sigma}}  
- \tilde{\lambda}_{\tilde{\sigma}}) \; = \; \sum_I \left(
\frac{w^I - 1}{m} \right) \; \rho_{I 0} \; e^{\phi^I 
+ 2 \tilde{\lambda} + 2 \tilde{\sigma}} + (m - 1) \; \left(
e^{2 \tilde{\lambda}} - 1 \right) 
\end{equation}
where $r^2 \; {\cal B} = \sum_i
\tilde{\lambda}^i_{\tilde{\sigma}} \; \lambda^i_{\tilde{\sigma}}
\;$. Thus, equations (\ref{w0rr}) -- (\ref{wsigma}) describe the
stars when the equations of state are linear as given in
equation (\ref{eoswi}) with $w^I_\Pi = w^I \;$.


\vspace{4ex} 

\centerline{\bf 4. Asymptotic solutions and perturbations}

\vspace{2ex}

We now study the solutions to equations (\ref{w0rr}) --
(\ref{wsigma}) in the limit $r \to \infty \;$. We consider the
ansatz 
\[
\tilde{\lambda}^0 = \tilde{\lambda}^0_0 + \tilde{s}^0 \;
\tilde{\sigma} + \tilde{u}^0 \; \; , \; \; \;
\lambda^i = \lambda^i_0 + s^i \; \tilde{\sigma} + u^i 
\; \; , \; \; \;
\tilde{\lambda} = \tilde{\lambda}_0 + \tilde{s} \;
\tilde{\sigma} + \tilde{u}
\]
where $(\tilde{\lambda}^0_0, \; \lambda^i_0, \;
\tilde{\lambda}_0)$ and $(\tilde{s}^0, \; s^i, \; \tilde{s})$
are constants, $\tilde{\lambda}^0_0$ and $\lambda^i_0$ can be
set to zero with no loss of generality, and $(\tilde{u}^0, \;
u^i, \; \tilde{u})$ are functions of $r \;$. In the limit $r \to
\infty \;$, the $\tilde{\lambda}_0$ and the $\tilde{\sigma}$
terms give the leading order asymptotic solutions, which are
referred to in \cite{cpaper} as singular solutions. The
functions $\tilde{u}$'s give the next order corrections, and
thus are the perturbations around the asymptotic solutions. In
this paper, we analyse the equations of motion to zeroth order
in the $\tilde{u}$'s to obtain the asymptotic solutions; and
analyse to the first order in the $\tilde{u}$'s to obtain
perturbation equations.

Set $\tilde{\lambda}^0_0 = \lambda^i_0 = 0 \;$ with no loss of
generality. Then
\[ 
\Lambda^c = S^c \; \tilde{\sigma} + U^c \; \; , \; \; \;
\tilde{\lambda}^i = \tilde{s}^i \; \tilde{\sigma} + \tilde{u}^i
\; \; , \; \; \; \phi^I = q^I \; \tilde{\sigma} + y^I
\] 
where
\[
S^c = \sum_i s^i \; \; , \; \; \;
U^c = \sum_i u^i 
\; \; \; ; \; \; \; \; 
\tilde{s}^i = s^i + \frac{S^c}{m} 
\; \; , \; \; \; 
\tilde{u}^i = u^i + \frac{U^c}{m} 
\]
and, as follows from equation (\ref{phii}), 
\begin{eqnarray} 
w^I \; q^I & = & - \; (1 + w^I) \; \tilde{s}^0 
- \sum_i \tilde{c}^{i I} \; s^i \label{qi} \\
w^I \; y^I & = & - \; (1 + w^I) \; \tilde{u}^0 
- \sum_i \tilde{c}^{i I} \; u^i \label{yi} \; \; . 
\end{eqnarray} 
Also, write $r^2 \; {\cal B} = \sum_i
\tilde{\lambda}^i_{\tilde{\sigma}} \; \lambda^i_{\tilde{\sigma}}
= {\cal B}_0 + 2 {\cal B}_1 + {\cal B}_2$ where
\[
{\cal B}_0 = \sum_i \tilde{s}^i \; s^i \; \; , \; \; \; 
{\cal B}_1 = \sum_i s^i \; \tilde{u}^i_{\tilde{\sigma}} 
= \sum_i \tilde{s}^i \; u^i_{\tilde{\sigma}} \; \; , \; \; \;
{\cal B}_2 = \sum_i \tilde{u}^i_{\tilde{\sigma}} \; 
u^i_{\tilde{\sigma}} \;\; .
\]

Using the above expressions, we now expand equations
(\ref{w0rr}) -- (\ref{wsigma}) to zeroth and first order in the
$\tilde{u}$'s. At zeroth order, equating the powers of $r$ gives
immediately
\begin{equation}
2 + q^I + 2 \tilde{s} \; = \; \tilde{s} \; = \; 0 \; \; . 
\end{equation}
Hence, 
\begin{eqnarray*}
r^2 (\rho_I e^{2 \lambda}) & = & \rho_{I 0} \; 
e^{ 2 \tilde{\lambda}_0 + y^I + 2 \tilde{u}} \; = \; 
R_I \; (1 + y^I + 2 \tilde{u} + \; \cdots \; ) \\
& & \\
e^{2 \tilde{\lambda}} - 1 & = & e^{2 \tilde{\lambda}_0 
+ 2 \tilde{u}} - 1 \; = \; (e^{2 \tilde{\lambda}_0} - 1) 
+ e^{2 \tilde{\lambda}_0} \; (2 \tilde{u} + \; \cdots \; ) \\
& & \\
\tilde{\chi}_{\tilde{\sigma}} - \tilde{\lambda}_{\tilde{\sigma}}
& = & \alpha + (\tilde{u}^0_{\tilde{\sigma}} 
- \tilde{u}_{\tilde{\sigma}}) \; \; \; , \; \; \; \;
\alpha \; = \; m - 1  + \tilde{s}^0 - \tilde{s} 
\end{eqnarray*}
where $R_I = \rho_{I 0} \; e^{2 \tilde{\lambda}_0} \;$. Then
equations (\ref{w0rr}) -- (\ref{yi}) give
\begin{eqnarray}
\alpha \; \tilde{s}^0 & = & \sum_I \tilde{c}^{0 I} \; R_I
\; \; \; , \; \; \; \; 
\alpha \; \tilde{s}^i \; = \; \sum_I \tilde{c}^{i I} \; R_I
\label{00i} \\
\tilde{s}^0 & = & \sum_I \frac{w^I}{m} \; R_I 
+ \frac{m - 1}{2} \; \left( e^{2 \tilde{\lambda}_0} - 1 \right)
+ \frac{{\cal B}_0}{2 m} \label{0quad} \\
\tilde{s}^0 - \tilde{s} & = & \sum_I 
\left(\frac{w^I - 1}{m} \right) \; R_I + (m - 1) \; 
\left( e^{2 \tilde{\lambda}_0} - 1 \right) \label{0sigma} \\
& & \nonumber \\
2 w^I & = & (1 + w^I) \; \tilde{s}^0 
+ \sum_i \tilde{c}^{i I} \; s^i \label{0i}
\end{eqnarray}
at zeroth order. And, they give
\begin{eqnarray}
\tilde{u}^0_{\tilde{\sigma} \tilde{\sigma}} 
+ \alpha \tilde{u}^0_{\tilde{\sigma}} 
+ \tilde{s}^0 \; (\tilde{u}^0_{\tilde{\sigma}} 
- \tilde{u}_{\tilde{\sigma}}) & = &
\sum_I \tilde{c}^{0 I} \; R_I \; 
(y^I + 2 \tilde{u}) \label{10} \\
\tilde{u}^i_{\tilde{\sigma} \tilde{\sigma}} 
+ \alpha \tilde{u}^i_{\tilde{\sigma}} 
+ \tilde{s}^i \; (\tilde{u}^0_{\tilde{\sigma}} -
\tilde{u}_{\tilde{\sigma}}) & = &
\sum_I \tilde{c}^{i I} \; R_I \; (y^I + 2 \tilde{u}) \label{1i}
\end{eqnarray}
\begin{eqnarray}
\tilde{u}^0_{\tilde{\sigma}} & = &
\sum_I \frac{w^I}{m} \; R_I \; (y^I + 2 \tilde{u}) 
+ (m - 1) \; e^{2 \tilde{\lambda}_0} \; \tilde{u}
+ \frac{{\cal B}_1}{m} \label{1quad} \\
& & \nonumber \\
\tilde{u}^0_{\tilde{\sigma}} - \tilde{u}_{\tilde{\sigma}} 
& = & \sum_I \left( \frac{w^I - 1}{m} \right) \; R_I \; 
(y^I + 2 \tilde{u}) + (m - 1) \; e^{2 \tilde{\lambda}_0} \; 
(2 \tilde{u}) \label{1sigma} \\
& & \nonumber \\
w^I \; y^I & = & - \; (1 + w^I) \; \tilde{u}^0 
- \sum_i \tilde{c}^{i I} \; u^i \label{1yi}
\end{eqnarray}
at first order. Note from equation (\ref{0sigma}) that
\[
\alpha \; = \; m - 1 + \tilde{s}^0 - \tilde{s} \; = \; 
\sum_I \left(\frac{w^I - 1}{m} \right) \; R_I + (m - 1) \; 
e^{2 \tilde{\lambda}_0} \; \; . 
\]

\vspace{1ex}

We now make a few general remarks, thereby also summarising very
briefly several results in the studies of stars which will be
used here. See \cite{cpaper} -- \cite{katz} for more details.

\vspace{1ex}

{\bf (1)} 
Given the equations of state, the constants $(w^I_\alpha, \;
\tilde{c}^{\alpha I})$ are known. The zeroth order equations are
simply algebraic equations for the constants $(\tilde{s}^\alpha,
\; R_I, \; e^{2 \tilde{\lambda}_0}) \;$. The first order
equations are linear differential equations for the
perturbations $\tilde{u}^\alpha \;$. They are of the type
encountered in standard `small oscillations' problems in
mechanics, and may be solved by standard techniques.
Nevertheless, solving these equations and obtaining a general
answer is algebraically quite involved. Hence, to proceed
further and to illustrate the nature of the results, we will
consider a few particular examples.

\vspace{1ex}

{\bf (2)} 
It follows from $\tilde{\lambda} = \tilde{\lambda}_0 +
\tilde{u}$ and from the definition $\frac{M(r)} {r^{m - 1}} = 1
- e^{- 2 \tilde{\lambda}}$ that, in the limit of large $r$,
\[
\frac{M(r)} {r^{m - 1}} \; \to \;
1 - e^{- 2 \tilde{\lambda}_0} 
\]
and that $\tilde{u}$ is the perturbation in the mass of the
star.

\vspace{1ex}

{\bf (3)} 
The size of the spherically symmetric star is defined as the
radius $r_*$ at which the pressure vanishes. The criteria for
finiteness of $r_*$ are given in \cite{rs}. The radius $r_*$ is
infinite for stars made up of perfect fluids with equation of
state $p = w \rho$ where $0 \le w \le 1 \;$. It is a standard
procedure then to construct composite configurations consisting
of perfect fluid cores and constant density crusts to render
$r_*$ finite, see \cite{cpaper} for example. Or, to confine the
system within a spherical box of radius $r_* \;$, see
\cite{pc07} for example.

OV equations are then solved numerically. The solutions contain
one free parameter, taken in \cite{pc07} to be $x_{ch} \propto
\sqrt{\rho_0} \; r_*$ which is a measure of central density. Let
$y_{ch} = \frac{M}{r_*^{m - 1}}$ which measures the mass of the
star. It is found from the numerical analyses that : {\bf (i)}
As $x_{ch}$ increases from zero to $\infty$, $\; y_{ch}$
increases from zero to a (first) maximum $y_1$ at $x_1$,
thereafter exhibits damped oscillations, asymptoting to a value
$y_s \;$. The asymptotic value $y_s$ is typically several
percent smaller than $y_1$ in magnitude. {\bf (ii)} The
behaviour for large values of $x_{ch}$ is described well by the
analytical asymptotic solutions and the perturbations around
it. {\bf (iii)} The solutions are unstable beyond the first
maximum which is at 
$(x_{ch},  \; y_{ch}) = (x_1, \; y_1) \;$. 

As a consequence, one has the following. For a given value of
central density, the radius $r_*$ must be $< r_{1 *}$ where
$r_{1 *}$ corresponds to $x_1$. The mass of the star must then
be less than $(y_1 \; r_{1 *}^{m - 1}) \;$. A more massive star
will be unstable and will collapse.

\vspace{1ex}

{\bf (4)} 
In \cite{pc07}, Chavanis studied static spherically symmetric
equilibrium configurations of stars in $D = m + 2$ dimensional
spacetime, $n_c = 0$ and ${\cal N} = 1$ in our notation. He
found that, for $m \ge m_{cr} (w) \sim 9$, $\; y_{ch}$ increases
monotonously from zero to $y_s$, effectively making $x_1$
infinite and $y_1 = y_s \;$, see Figures 20 -- 23 in
\cite{pc07}. The perturbations around the corresponding
analytical asymptotic solutions exhibit a monotonous behaviour
with no oscillations.

As a consequence, one has the following. For $m \ge 9$, $\; x_1$
is effectively infinite. Then $r_{1 *}$ is infinite which makes
the upper limit $(y_s \; r_{1 *}^{m - 1})$ on the mass of the
star also infinite. Hence, a star can be arbitrarily massive and
stable when $m \ge 9 \;$.


\vspace{4ex}

\centerline{\bf Example (1) : $\; \; \; {\mathbf \tilde{c}^{i I}
= 0 \; \; \Longrightarrow \; \; s^i = 0 \;}$}

\vspace{2ex}

We first consider an example where the corresponding stars are
effectively same as those in \cite{pc07}. Consider the case
where $w^I_i = \frac{(m + 1) w^I - 1}{m} \;$. Then $\tilde{c}^{i
I} = 0$ from which it follows that $s^i = 0 = S^c = {\cal B}_0 =
{\cal B}_1 \;$, so we omit the tilde's on $s$'s. We then get
$s^0 = \frac{2 w^I}{1 + w^I}$ which implies that $w^I = w$ for
all $I \;$ \cite{fn1}. Then
\[
\alpha = m - 1 + \frac{2 w}{1 + w} \; \; , \; \; \; 
\tilde{c}^{0 I} = \tilde{c}^0 = \frac{m - 1 + (m + 1) w}{m} 
\; \; .
\]
Note that $(1 + w) \alpha = m \tilde{c}^0 \;$. The zeroth order
equations (\ref{00i}) -- (\ref{0sigma}) give
\[
\alpha s^0 \; = \; \tilde{c}^0 \sum_I R_I 
\; \; \; \Longrightarrow \; \; \; 
R = \sum_I R_I = \frac{2 m w}{(1 + w)^2}  
\]
\[
s^0 \; = \; \frac{w}{m} \; R + \frac{m - 1}{2} \; 
\left( e^{2 \tilde{\lambda}_0} - 1 \right) 
\; \; \; \Longrightarrow \; \; \; 
e^{2 \tilde{\lambda}_0} = \frac{{\cal D}}{(m - 1) (1 + w)^2}
\]
where ${\cal D} = (m - 1) (1 + w)^2 + 4 w \;$. The remaining
relation
\[
s^0 = \frac{w - 1}{m} \; R + (m - 1) \; 
\left( e^{2 \tilde{\lambda}_0} - 1 \right)
\]
is now satisfied identically. 

At first order, since $\tilde{c}^{i I} = 0$ and $w^I = w$, we
have
\[
w \; y^I \; = \; - \; (1 + w) \; \tilde{u}^0 
\; \; \; , \; \; \;\; 
\tilde{u}^i_{\tilde{\sigma} \tilde{\sigma}} + \alpha
\tilde{u}^i_{\tilde{\sigma}} \; = \; 0 \; \; . 
\]
Thus, $y^I = y$ is independent of $I$ and $\tilde{u}^i \;$.
Equations (\ref{10}), (\ref{1quad}), and (\ref{1sigma}) now give
\[
\tilde{u}^0_{\tilde{\sigma} \tilde{\sigma}} 
+ \alpha \tilde{u}^0_{\tilde{\sigma}} 
+ \tilde{s}^0 \; (\tilde{u}^0_{\tilde{\sigma}} 
- \tilde{u}_{\tilde{\sigma}}) \; = \; 
\tilde{c}^0 \; R \; (y + 2 \tilde{u}) \;\; , 
\]

\begin{eqnarray*}
\tilde{u}^0_{\tilde{\sigma}} & = & \frac{w}{m} \; R \; 
(y + 2 \tilde{u}) + (m - 1) \; e^{2 \tilde{\lambda}_0} \;
\tilde{u} \\ 
& & \\
\tilde{u}^0_{\tilde{\sigma}} - \tilde{u}_{\tilde{\sigma}} 
& = & \frac{w - 1}{m} \; R \; (y + 2 \tilde{u}) + (m - 1) \;
e^{2 \tilde{\lambda}_0} \; (2 \tilde{u}) \; \; , 
\end{eqnarray*}
from which it follows that
\begin{eqnarray*}
\tilde{u}^0_{\tilde{\sigma}} & = & - \; \frac{2 w}{1 + w} \; 
\tilde{u}^0 
+ \left( m - 1 + \frac{4 w}{1 + w} \right) \; \tilde{u} \\ 
\tilde{u}_{\tilde{\sigma}} & = & - \; \frac{2}{1 + w} \; 
\tilde{u}^0 - (m - 1) \; \tilde{u} \\ 
\tilde{u}^0_{\tilde{\sigma}} & = & w \tilde{u}_{\tilde{\sigma}} 
+ \frac{{\cal D}}{1 + w} \; \tilde{u} \; \; .
\end{eqnarray*}

The $\tilde{u}^0_{\tilde{\sigma}}$ and
$\tilde{u}_{\tilde{\sigma}}$ equations above are of the type
\begin{equation}\label{ABPQ}
A x' + B y' = a x + b y \; \; , \; \; \; 
P x' + Q y' = p x + q y 
\end{equation}
in an obvious notation. Solving, for example, for $(x', x)$ in
terms of $(y', y)$ gives $x' = a_1 y' + a_2 y \;$ and $x = a_3
y' + a_4 y \;$; one then gets $a_3 y'' + a_4 y' = x' = a_1 y' +
a_2 y$ which gives an equation for $y'' \;$. An equation for
$x'' \;$ also follows similarly. One can show in this way that
both $x''$ and $y''$ obey the same equation, namely
\begin{equation}\label{*''}
(P B - Q A) \; (*)'' - (P b + p B - Q a - q A) \; (*)' 
+ (p b - q a) \; (*) \; = \; 0
\end{equation}
where $(*) = x, \; y \;$. Applying this formula, it follows
straightforwardly that $\tilde{u}^0$ and $\tilde{u}$ obey the
same equation, namely  
\[
\tilde{u}^0_{\tilde{\sigma} \tilde{\sigma}} 
+ \alpha \tilde{u}^0_{\tilde{\sigma}} + 
\frac{2 {\cal D}}{(1 + w)^2} \; \tilde{u}^0 \; = \; 0 
\]
\[
\tilde{u}_{\tilde{\sigma} \tilde{\sigma}} 
+ \alpha \tilde{u}_{\tilde{\sigma}} + 
\frac{2 {\cal D}}{(1 + w)^2} \; \tilde{u} \; = \; 0 \; \; .
\]
Except for the presence of $\tilde{u}^i$s obeying the equation
$\tilde{u}^i_{\tilde{\sigma} \tilde{\sigma}} + \alpha
\tilde{u}^i_{\tilde{\sigma}} = 0 \;$, the results in this
example are same as the corresponding ones of \cite{pc07} for
which $n_c = 0 \;$.  This is because the choice of $w^I_i$,
leading to $s^i = 0$, decouples the effects of compact space.


\vspace{4ex}

\centerline{\bf Example (2) : $\; \; \; {\mathbf \tilde{c}^{i I}
\ne 0 \; \; , \; \; \; {\cal N} = 1 \;}$}

\vspace{2ex}

Consider an example where $\tilde{c}^{i I} \ne 0$ but ${\cal N}
= 1 \;$. Using $\tilde{s} = 0$, $\; q^I = - 2 \;$, and omitting
the $I-$scripts, the zeroth order equations (\ref{qi}), (\ref{00i}) --
(\ref{0sigma}) give
\begin{eqnarray*}
2 \; w & = & (1 + w) \; \tilde{s}^0 
+ \sum_i \tilde{c}^i \; s^i \\
\alpha \; \tilde{s}^0 & = & \tilde{c}^0 \; R 
\; \; \; , \; \; \; \; 
\alpha \; \tilde{s}^i \; = \; \tilde{c}^i \; R \\
& & \\
\tilde{s}^0 & = & \frac{w}{m} \; R
+ \frac{m - 1}{2} \; \left( e^{2 \tilde{\lambda}_0} - 1 \right)
+ \frac{{\cal B}_0}{2 m} \\ & & \\
\tilde{s}^0 & = & \frac{w - 1}{m} \; R 
+ (m - 1) \; \left( e^{2 \tilde{\lambda}_0} - 1 \right) \\
& & \\
\Longrightarrow \; \; \; \; 
\alpha & = & m - 1 + \tilde{s}^0 \; = \; \frac{w - 1}{m} \; R 
+ (m - 1) \; e^{2 \tilde{\lambda}_0}
\end{eqnarray*}
where $ {\cal B}_0 = \sum_i \tilde{s}^i \; s^i \;$. We now get
\[
\left( \tilde{s}^i, \; s^i, \; R \right) \; = \;
\left( \tilde{c}^i, \; c^i, \; \alpha \right) \; 
\frac {\tilde{s}^0} {\tilde{c}^0} \; \; , 
\]
using which we obtain 
\[
\tilde{s}^0 = \frac{2 w}{(1 + w) (1 + \gamma)}
\; \; \; , \; \; \; \; 
\gamma \; = \; \frac{\sum_i \tilde{c}^i \; c^i} {(1 + w)
\tilde{c}^0} \; \; .
\]
Then $\tilde{s}^i$ and $\alpha = m - 1 + \tilde{s}^0$ follow,
then $R$, then $e^{2 \tilde{\lambda}_0}$ from, for example, the
last of the zeroth order equations:
\[
(m - 1) \; e^{2 \tilde{\lambda}_0} \; = \; \alpha + \frac{1 -
w}{m} \; R \; = \; \frac{\alpha}{1 + \gamma} \; \left( \frac
{{\cal D}} {m \tilde{c}^0 (1 + w)} + \gamma \right) \; \; . 
\]
Also, $ {\cal B}_0 = \sum_i \tilde{s}^i s^i = \gamma (1 + w) \;
\frac {(\tilde{s}^0)^2} {\tilde{c}^0} \;$. All the results of
Example (1) will follow from the present ones upon setting
$\gamma = 0 \;$.

The first order equations (\ref{yi}), (\ref{10}) --
(\ref{1sigma}) give 
\begin{eqnarray*}
w \; y + (1 + w) \; \tilde{u}^0 + \sum_i \tilde{c}^i \; u^i 
& = & 0 \\
& & \\
\tilde{u}^0_{\tilde{\sigma} \tilde{\sigma}} + \alpha
\tilde{u}^0_{\tilde{\sigma}} + \tilde{s}^0 \;
(\tilde{u}^0_{\tilde{\sigma}} - \tilde{u}_{\tilde{\sigma}}) 
& = & \tilde{c}^0 \; R \; (y + 2 \tilde{u}) \\
& & \\
\tilde{u}^i_{\tilde{\sigma} \tilde{\sigma}} + \alpha
\tilde{u}^i_{\tilde{\sigma}} + \tilde{s}^i \;
(\tilde{u}^0_{\tilde{\sigma}} - \tilde{u}_{\tilde{\sigma}}) 
& = & \tilde{c}^i \; R \; (y + 2 \tilde{u}) 
\end{eqnarray*}
\begin{eqnarray*}
\tilde{u}^0_{\tilde{\sigma}} & = & \frac{w}{m} \; R \; 
(y + 2 \tilde{u}) + (m - 1) \; e^{2 \tilde{\lambda}_0} \;
\tilde{u} + \frac{{\cal B}_1}{m} \\
& & \\
\tilde{u}^0_{\tilde{\sigma}} - \tilde{u}_{\tilde{\sigma}} 
& = & \frac{w - 1}{m} \; R \; (y + 2 \tilde{u}) 
+ (m - 1) \; e^{2 \tilde{\lambda}_0} \; (2 \tilde{u})
\end{eqnarray*}
where ${\cal B}_1 = \sum_i s^i \tilde{u}^i_{\tilde{\sigma}} \;$.

\vspace{2ex}

It can now be seen from the $\tilde{u}^0_{\tilde{\sigma}
\tilde{\sigma}}$ and $\tilde{u}^i_{\tilde{\sigma}
\tilde{\sigma}}$ equations, and from the zeroth order results
for $\tilde{s}^i$, that
\[
F^i_{\tilde{\sigma} \tilde{\sigma}} + \alpha \; 
F^i_{\tilde{\sigma}}  \; = \; 0 
\; \; \;, \; \; \; \; 
F^i = \tilde{u}^i - \tilde{c}^i \; \frac {\tilde{u}^0}
{\tilde{c}^0} \; \; . 
\]
Although $F^i$ admit the general solutions $F^i = F^i_0 + F^i_1
\; e^{- \alpha \tilde{\sigma}} \;$, we will set the integration
constants $F^i_{0, 1}$ to zero and thus take $\tilde{u}^i =
\tilde{c}^i \; \frac {\tilde{u}^0} {\tilde{c}^0} \;$. It is now
straightforward to obtain $y$ and ${\cal B}_1 \;$. They are
given by
\[
\tilde{s}^0 \; y = - \; 2 \; \tilde{u}^0 
\; \; , \; \; \; 
{\cal B}_1 = \gamma  (1 + w) \tilde{s}^0 \; 
\frac {\tilde{u}^0} {\tilde{c}^0} \; \; .  
\]
Using the above expression for $y$, and after some algebra, we
now get
\[
\tilde{u}^0_{\tilde{\sigma} \tilde{\sigma}} + \alpha
\tilde{u}^0_{\tilde{\sigma}} + \frac{2 \; \alpha}{1 + \gamma}
\left( \frac {{\cal D}} {m \tilde{c}^0 (1 + w)} + \gamma \right)
\tilde{u}^0 \; = \; 0 
\]

\begin{eqnarray*}
\left( 1 - \frac {\gamma (1 + w) \tilde{s}^0} {m \tilde{c}^0}
\right) \; \tilde{u}^0_{\tilde{\sigma}} & = & 
- \; \frac {2 \alpha w} {m \tilde{c}^0} \; \tilde{u}^0 
+ \alpha \left( 1 + \frac {(1 + w) \tilde{s}^0} {m \tilde{c}^0}
\right) \tilde{u} \\ & & \\ 
\tilde{u}^0_{\tilde{\sigma}} - \tilde{u}_{\tilde{\sigma}} 
& = & \frac {2 \alpha (1 - w)} {m \tilde{c}^0} \; \tilde{u}^0
+ 2 \alpha \tilde{u} \; \; .
\end{eqnarray*}
The $\tilde{u}^0_{\tilde{\sigma}}$ and
$\tilde{u}_{\tilde{\sigma}}$ equations above are of the type
given in equation (\ref{ABPQ}). Hence, applying the consequent
formula (\ref{*''}) and after some algebra, one can show that
both $\tilde{u}^0_{\tilde{\sigma} \tilde{\sigma}}$ and
$\tilde{u}_{\tilde{\sigma} \tilde{\sigma}}$ obey the same
equation, given above for $\tilde{u}^0_{\tilde{\sigma}
\tilde{\sigma}} \;$.

The results of this example provide a generalisation of the
corresponding ones of \cite{pc07}. In this example, ${\cal N} =
1$ and the effects of compact space appear through a single
parameter $\gamma \;$.


\vspace{4ex}

\centerline{\bf Analysis of perturbation equation} 

\vspace{2ex}

Consider the equation for the perturbations
\[
(*)'' + \alpha \; (*)' + 2 \; \alpha \left( 1 
+ \frac{(1 - w) {\tilde{s}^0}} {m \tilde{c}^0} \right) \; (*)
\; = \; 0
\]
where $(*) = \tilde{u}^0$ or $\tilde{u}$, and we have used
\[
\frac{\alpha}{1 + \gamma} \; \left( \frac {{\cal D}} 
{m \tilde{c}^0 (1 + w)} + \gamma \right) \; = \; \alpha 
+ \frac{1 - w}{m} \; R
\; \; \; , \; \;\; \; 
R = \frac {\alpha \tilde{s}^0} {\tilde{c}^0} \; \; . 
\]
Then the solutions are given by $(*) \sim e^{k \;
\tilde{\sigma}}$ where 
\[
k = \frac {- \alpha \pm \sqrt{\Delta}} {2} 
\; \; \; , \; \;\; \; 
\Delta = \alpha^2 - 8 \; \alpha \left( 1 + \frac{(1 - w)
{\tilde{s}^0}} {m \tilde{c}^0} \right) \; \; . 
\]
If $\Delta$ is negative then $k$ is complex and the solutions
exhibit damped oscillations. Such oscillations imply an
instability towards a collapse of a sufficiently massive star.
If $\Delta$ is zero or positive then $k$ is real and the
solutions are exponentially damped but with no oscillations.
Such non oscillating, monotonous behaviour implies stability
\cite{pc07}. The fluctuations of $\tilde{u}$ are related to the
fluctuations of the mass function $M(r)$ and, hence, a negative
$\Delta$ denotes an instability towards a collapse of a
sufficiently massive star whereas a zero or a positive $\Delta$
denotes stability.

In order to analyse the sign of $\Delta$, we write it as
\begin{eqnarray*}
\Delta & = & \kappa_1 \left\{ \alpha \; m \tilde{c}^0 
- 8 \; (m \tilde{c}^0 + (1 - w) \tilde{s}^0) \right\} \\
& & \\
& = & \kappa_1 \left\{ (m - 9) \; m \tilde{c}^0 + \tilde{s}^0 \;
(m \tilde{c}^0 - 8 (1 - w)) \right\} \\
& & \\
& = & \kappa_2 \left\{ (m - 9) (1 + w) (1 + \gamma) \; 
m \tilde{c}^0 + 2 w \; (m \tilde{c}^0 - 8 (1 - w)) \right \} \\
& & \\
& = & \kappa_2 \left\{ b_2 w^2 + b_1 w + b_0 \right\}
\end{eqnarray*}
where $\kappa_1 = \frac{\alpha}{m \tilde{c}^0} \;$, $\; \kappa_2
= \frac{\kappa_1}{(1 + w) (1 + \gamma)} \;$, and
\begin{eqnarray*}
b_2 & = & (m - 3)^2 + \gamma (m + 1) (m - 9) \\
b_1 & = & 2 (m - 9) (m + 1 + \gamma m) \\
b_0 & = & (m - 1) (m - 9) (1 + \gamma) \; \; .
\end{eqnarray*}
We take $c^i$ to not all vanish, then the sum $\sum_i
\tilde{c}^i c^i > 0 \;$. Assume that $w > \frac{1 - m}{1 + m}
\;$. It is, indeed, physically natural that $0 \le w \le 1 \;$.
It then follows that $(1 + w, \; \tilde{c}^0, \; \gamma, \;
\tilde{s}^0, \; \alpha, \; \kappa_1, \; \kappa_2)$ are all
positive and, hence,
\[
Sgn \; \Delta \; = \; Sgn \; (b_2 w^2 + b_1 w + b_0) \; \; .
\]

We now analyse the sign of $\Delta \;$. Write $b_2 w^2 + b_1 w +
b_0 = Q_0 + Q_1$ where
\[
Q_0 \; = \; (m - 3)^2 w^2 + 2 (m - 9) (m + 1) w 
+ (m - 1) (m - 9)
\]
and 
\[
Q_1 \; = \; \gamma (m - 9) (1 + w) \; m \tilde{c}^0 \; = \; 
m (m - 9) \; \sum_i \tilde{c}^i c^i \; \; .
\]
Note that the polynomial $Q_0$ is same as that analysed in
\cite{pc07}; that the discriminant of $Q_0$ is $128 \; m \; (9 -
m) \;$; that $\sum_i \tilde{c}^i c^i > 0 \;$; and that the
polynomial $Q_1$ may be thought of as arising due to the effects
of compact space. Now consider $Q_0 + Q_1$ for different ranges
of $m \;$.

\vspace{1ex}

\noindent
${\mathbf m > 9}$ : In this case, $Q_1$ is positive. The
discriminant of $Q_0$ is negative, hence $Q_0$ is positive for
all $w \;$. Therefore, $\Delta$ is positive.

\vspace{1ex}

\noindent
${\mathbf m = 9}$ : In this case, $Q_0 = 36 w^2$ and $Q_1 = 0
\;$. Hence, $\Delta$ is zero or positive. 

\vspace{1ex}

\noindent
${\mathbf m < 9}$ : In this case, $Q_1 < 0 \;$. Consider $Q_0$.
It is simple to show \cite{pc07} that $Q_0$ is negative for $0
\le w \le 1 \;$. Hence, $\Delta$ is negative for $0 \le w \le 1
\;$.

Thus, it follows that the perturbations exhibit monotonous and
non oscillating behaviour for $m \ge 9$ for any value of $w \;$.
Hence, a star can be arbitrarily massive and stable in these
cases. For $m < 9 \;$, and for the physically relevant range $0
\le w \le 1 \;$, the perturbations exhibit damped oscillations.
Therefore there will be instability and, hence, sufficiently
massive stars will collapse.


\vspace{4ex}

\centerline{\bf 5. M theory branes : U duality relation among
$(p_{\alpha I}, \Pi_I) \;$}

\vspace{2ex}

M theory has U duality symmetries. Incorporating these
symmetries leads to a relation among the components $(p_{\alpha
I}, \; \Pi_I)$ of the energy momentum tensor for the $I^{th}$
stack of branes. We will explain this relation below. Then, we
apply the general results of sections {\bf 3} and {\bf 4} to M
theory branes.

We use the U duality symmetries here the same way as in our
earlier works \cite{k07}. We note that, for a given ${\cal
N}$, different intersecting brane configurations of M theory can
be related to each other by suitable U duality operations --
namely, by suitable dimensional reduction and uplifting to and
from type IIA string theory, and T and S dualities in type IIA/B
string theories. For a metric of the form given in equation
(\ref{ds}), such an operation leads to relations among
$\lambda^\alpha$ which in turn, through their equations of
motion, imply relations among $(p_\alpha, \Pi) \;$. Although
only time dependent cases in early universe were studied in
\cite{k07}, the U duality details of these works carry over
and are applicable to the static cases also with only a few
minor changes. We present only the main results here, see
\cite{k07} for details.

It can be shown that the relations among $(p_\alpha, \Pi)$ for
intersecting branes of M theory, obtained by applying U duality
operations as described above, are all satisfied if the
individual $(p_{\alpha I}, \Pi_I)$ obey the relation
\begin{equation}\label{Ups}
p_{\parallel I} \; = \; \Pi_I + p_{0 I} + p_{\perp I} 
+ m \; (p_{\Omega I} - p_{\perp I}) 
\end{equation} 
where $p_{\parallel I}$ and $p_{\perp I}$ are the pressures
along the directions that are parallel and transverse to the
worldvolume of the $I^{th}$ stack of branes. The above relation
is a consequence of U duality symmetries and, therefore, must
always be valid independent of the details of the equations of
state. We further take $p_{\Omega I} = p_{\perp I}$ which is
natural since the sphere directions are transverse to the
branes. Furthermore, in the case of early universe studied
previously as well as in the case of stars that are being
studied here, the constituent matter components satisfy $T^r_{\;
\; r (I)} = T^a_{\; \; a (I)}$, thus $\Pi_I = p_{a I} \;$. Using
$p_{0 I} = - \rho_I$ and $p_{a I} = p_{\Omega I} = p_{\perp I} =
p_I$, equation (\ref{Ups}) then becomes $p_{\parallel I} = -
\rho_I + 2 p_I \;$ \cite{cm, k07}.

Consider the case where the equations of state are linear, as
given in equations (\ref{eoswi}). For M theory branes, let
$p_{\parallel I} = w^I_\parallel \; \rho_I$ and $p_{\perp I} =
w^I_\perp \; \rho_I \;$. Taking $w^I_\perp = w^I_\Omega = w^I$,
and then $w^I_\Pi = w^I_a$ for stars, the U duality relation in
equation (\ref{Ups}) then gives
\[
w^I_\parallel = w^I_\Pi - 1 + w^I = - 1 + 2 w^I \; \; . 
\]
For intersecting branes in M theory, these relations further
lead to an elegant structure which is shown in Appendix {\bf C}.

Consider the constants $\tilde{c}^{i I}$ defined in equation
(\ref{tldcalpha}). For intersecting branes in M theory, $w^I_i =
w^I_\parallel$ if $\; i \in \; \parallel_I \;$, namely if $x^i$
is a worldvolume coordinate of the $I^{th}$ stack of branes, and
$w^I_i = w^I_\perp$ otherwise. Similarly, let $\tilde{c}^{i I} =
\tilde{c}^{\parallel I} \;$ if $\; i \in \; \parallel_I $ and
$\tilde{c}^{i I} = \tilde{c}^{\perp I} $ otherwise. Then, using
the U duality relation $w^I_\parallel = - 1 + 2 w^I$, it follows
that
\begin{equation}\label{uc}
\tilde{c}^{\perp I} = \frac{w^I - 1}{m} 
\; \; , \; \; \;
\tilde{c}^{\parallel I} = \tilde{c}^{\perp I} + 1 - w^I 
=  (1 - m) \; \tilde{c}^{\perp I} \; \; . 
\end{equation}
Hence, for any $a^i$ with $a^c = \sum_i a^i$, it follows that
\begin{equation}\label{cai}
\sum_i \tilde{c}^{i I} \; a^i \; = \; 
(1 - w^I) \; \left( \sum_{i \in \parallel_I} a^i 
- \frac{a^c}{m} \right) \; \; . 
\end{equation}

With $w^I_\alpha$, and thus $\tilde{c}^{\alpha I}$, specified
for intersecting branes, it is now a straightforward exercise to
apply the general results of sections {\bf 3} and {\bf 4} to
stars made up of intersecting branes in M theory. In the
following, we present two examples: the ${\cal N} = 1$ case of
$M2$ and $M5$ brane stars, and the ${\cal N} = 4$ case of $2 2'
5 5'$ intersecting brane configuration.

Other intersecting brane configurations can be analysed by same
method as shown for the ${\cal N} = 4$ case. The results for the
${\cal N} = 3$ case of $2 2' 2''$ intersecting brane
configuration are almost identical to those of the ${\cal N} =
4$ case.


\vspace{4ex}

\centerline{\bf Example (3) : $\; {\mathbf {\cal N} = 1} \;$
case -- $\; {\mathbf M 2} \;$ and $\; {\mathbf M 5} \;$ branes}

\vspace{2ex} 

Consider stars made up of $M2$ branes or $M5$ branes. Then
${\cal N} = 1$, $\; D = 11$, $\; n_c = 2$ or $5$, $\; m = 7$ or
$4$, and $\tilde{c}^i = \tilde{c}^\parallel = \frac{(m - 1) (1 -
w)}{m} \;$. Hence,
\[
\sum_i \tilde{c}^i c^i = 
\frac{m \; n_c \; (\tilde{c}^\parallel)^2}{n_c + m} = 
\frac{n_c \; (m - 1)^2 \; (1 - w)^2}{m \; (n_c + m)} \; \; , 
\]
which $= \frac{8}{7} \; (1 - w)^2$ for $M2$ branes and $=
\frac{5}{4} \; (1 - w)^2$ for $M5$ branes.

This is a special case of Example (2) described in section 4.
Therefore, the analysis and the results of Example (2) apply
straightforwardly. Since $m < 9 \;$, the perturbations exhibit
damped oscillations for $0 \le w \le 1 \;$. Therefore there will
be instability and, hence, sufficiently massive $M2$ or $M5$
brane stars will collapse.


\vspace{4ex}

\centerline{\bf Example (4) : $\; {\mathbf {\cal N} = 4} \;$
case -- $\; {\mathbf 2 2' 5 5'} \;$ configuration}

\vspace{2ex} 

For the BPS intersecting brane configurations of the M theory,
we have
\[
n_c + m = 9 \; \; , \; \; \; w^I_\perp = w^I_\Omega = w^I
\; \; , \; \; \; w^I_\parallel = - 1 + 2 w^I \; \; .
\]
Also $\tilde{c}^{i I} = \tilde{c}^{\parallel I} \;$ if $\; i \in
\; \parallel_I $ and $\tilde{c}^{i I} = \tilde{c}^{\perp I} $
otherwise. From the relation between $w^I_\parallel$ and
$w^I_\perp$, it follows that
\[
\tilde{c}^{\perp I} = \frac{w^I - 1}{m} 
\; \; , \; \; \;
\tilde{c}^{\parallel I} = 1 - w^I + \tilde{c}^{\perp I} 
=  (1 - m) \; \tilde{c}^{\perp I} \; \; . 
\]

In the $2 2' 5 5' \; : \; (12, 34, 13567, 24567)$ configuration
of the M theory, there are two stacks each of two branes and
five branes whose spatial worldvolume directions are indicated.
For this configuration, we have
\[
{\cal N} = 4 \; \; , \; \;\; n_c = 7 \; \; , \; \;\; 
m = 2 \; \; , \; \;\;
\tilde{c}^{\parallel I} = - \; \tilde{c}^{\perp I} = 
\frac{1 - w^I}{2} \; \; .
\]
Using $\tilde{c}^{\parallel I} = 1 - w^I + \tilde{c}^{\perp I}$
and then $\tilde{c}^{\perp I} = \frac{w^I - 1}{m}$, we have
\begin{eqnarray*}
\sum_i \tilde{c}^{i I} \; s^i & = & (1 - w^I) \; 
\sum_{i \in \parallel_I} s^i + \tilde{c}^{\perp I} \; S^c 
\; = \; (1 - w^I) \; \left( \sum_{i \in \parallel_I} s^i 
- \frac{S^c}{m} \right) \\
\sum_i \tilde{c}^{i I} \; u^i & = & (1 - w^I) \; 
\sum_{i \in \parallel_I} u^i + \tilde{c}^{\perp I} \; U^c 
\; = \; (1 - w^I) \; \left( \sum_{i \in \parallel_I} u^i 
- \frac{U^c}{m} \right)  \; \; .
\end{eqnarray*}

Consider sums of the type $\sum_I \tilde{c}^{i I} \; X_I \;$.
Denoting the $X_I$'s for the $2 2' 5 5'$ configuration as $(X_2,
\; X_{2'}, \; X_5, \; X_{5'})$ and using $\tilde{c}^\parallel =
- \; \tilde{c}^\perp$, we get
\begin{eqnarray} 
\sum_I \tilde{c}^{1 I} \; X_I & = & \tilde{c}^\parallel \; 
\left( X_2 - X_{2'} + X_5 - X_{5'} \right) \nonumber \\
\sum_I \tilde{c}^{2 I} \; X_I & = & \tilde{c}^\parallel \; 
\left( X_2 - X_{2'} - X_5 + X_{5'} \right) \nonumber \\
\sum_I \tilde{c}^{3 I} \; X_I & = & \tilde{c}^\parallel \; 
\left( - X_2 + X_{2'} + X_5 - X_{5'} \right) \nonumber \\
\sum_I \tilde{c}^{4 I} \; X_I & = & \tilde{c}^\parallel \; 
\left( - X_2 + X_{2'} - X_5 + X_{5'} \right) \nonumber \\
\sum_I \tilde{c}^{5, 6, 7 \; \; I} \; X_I & = &
\tilde{c}^\parallel \;
\left( - X_2 - X_{2'} + X_5 + X_{5'} \right) \; \; .
\label{ciXi}
\end{eqnarray}

\vspace{2ex}

\noindent{\bf Zeroth order : }
At zeroth order, we have $\tilde{s} = 0$, $\; q^I = - 2$ and,
using the above expression for $\sum_i \tilde{c}^{i I} s^i$, we
get from equation (\ref{qi}) that
\[
\sum_{i \in \parallel_I} s^i \; = \; \frac{2 w^I - (1 + w^I)
\; \tilde{s}^0} {1 - w^I} \; + \; \frac{S^c}{m} \; \; .
\]
We now consider the case where $w^I = w$ for all $I$, and omit
the $I$ superscripts on $\tilde{c}^{0 I}$ and $w$'s. Note that
$\tilde{c}^{i I}$ will depend on $I$ since $\tilde{c}^{i I} =
\tilde{c}^{\parallel I} \;$ if $\; i \in \; \parallel_I $ and
$\tilde{c}^{i I} = \tilde{c}^{\perp I} \;$. Since $w^I = w$ for
all $I$, the above expression implies that the sum $\sum_{i \in
\parallel_I} s^i$ must be same for all $I \;$. Thus, for the $2
2' 5 5' \; : \; (12, 34, 13567, 24567)$ configuration, it
follows that
\[
s^1 + s^2 = s^3 + s^4 = s^1 + s^3 + s^5 + s^6 + s^7 
= s^2 + s^4 + s^5 + s^6 + s^7
\]
\[
\Longrightarrow \; \; \; 
s^3 = s^2 \;\; , \; \; \;  
s^4 = s^1 \;\; , \; \; \;  
s^5 + s^6 + s^7 = 0 \;\; , \; \; \;  
S^c = 2 \; (s^1 + s^2) \; \; .
\]
It can then be shown that $\tilde{s}^i = s^i + \frac{S^c}{m} \;$
satisfy the relations
\[
\tilde{s}^3 = \tilde{s}^2 \;\; , \; \; \;  
\tilde{s}^4 = \tilde{s}^1 \;\; , \; \; \;  
\tilde{s}^5 + \tilde{s}^6 + \tilde{s}^7 = \tilde{s}^1 +
\tilde{s}^2 \; \; .
\]

Consider now equation (\ref{00i}) for $\tilde{s}^i \;$: $\alpha
\; \tilde{s}^i = \sum_I \tilde{c}^{i I} R_I \;$. Denoting the
$R_I$'s for the $2 2' 5 5'$ configuration as $(R_2, \; R_{2'},
\; R_5, \; R_{5'})$ and using $\tilde{c}^\parallel = - \;
\tilde{c}^\perp$, we get
\begin{eqnarray*}
\alpha \; \tilde{s}^1 \; = \; 
\sum_I \tilde{c}^{1 I} R_I & = & \tilde{c}^\parallel \; 
\left( R_2 - R_{2'} + R_5 - R_{5'} \right) \\
\alpha \; \tilde{s}^2 \; = \; 
\sum_I \tilde{c}^{2 I} R_I & = & \tilde{c}^\parallel \; 
\left( R_2 - R_{2'} - R_5 + R_{5'} \right) \\
\alpha \; \tilde{s}^3 \; = \; 
\sum_I \tilde{c}^{3 I} R_I & = & \tilde{c}^\parallel \; 
\left( - R_2 + R_{2'} + R_5 - R_{5'} \right) \\
\alpha \; \tilde{s}^4 \; = \; 
\sum_I \tilde{c}^{4 I} R_I & = & \tilde{c}^\parallel \; 
\left( - R_2 + R_{2'} - R_5 + R_{5'} \right) \\
\alpha \; \tilde{s}^{5,6,7} \; = \; 
\sum_I \tilde{c}^{5,6,7 \; \; I} \; R_I & = &
\tilde{c}^\parallel \;
\left( - R_2 - R_{2'} + R_5 + R_{5'} \right) \; \; . 
\end{eqnarray*}
The three relations on the $\tilde{s}^i$ can now be seen to
imply that
\[
R_2 = R_{2'} = R_5 = R_{5'} = \frac{R}{4} 
\; \; , \; \; \; R = \sum_I R_I \; \; , 
\]
\[
\Longrightarrow \; \; \; 
\tilde{s}^i = 0 \; \; \; \Longrightarrow \; \; \; 
s^i = S^c = {\cal B}_0 = {\cal B}_1 = 0 \; \; . 
\]
Since $S^c = 0$, we omit the tilde's on $s$'s. Using equations
(\ref{00i}) -- (\ref{0sigma}), we then get the same zeroth order
results as in Example (1), namely
\[
s^0 = \frac{2 w}{1 + w} \; \; , \; \; \; 
\tilde{c}^{0 I} = \tilde{c}^0 = 
\frac{m - 1 + (m + 1) w}{m} \; \; , 
\]
\[
R = \sum_I R_I = \frac{2 m w}{(1 + w)^2} \; \; , \; \; \; 
(m - 1) \; e^{2 \lambda_0} = \frac{{\cal D}}{(1 + w)^2} 
\]
where ${\cal D} = (m - 1) (1 + w)^2 + 4 w$ and $m = 2 \;$. 


\vspace{2ex}

\noindent{\bf First order : } 
Consider now the equations at the first order in the
$\tilde{u}$'s for the case where $w^I = w$ for all $I \;$. Then
$(w^I, \; \tilde{c}^{0 I}, \; R_I)$ do not depend on $I$ and we
omit the $I-$scripts on them. Noting that $\tilde{s}^i = {\cal
B}_1 = 0$ and $R_I = \frac{R}{4}$ now, the first order equations
(\ref{10}) -- (\ref{1sigma}) for $y^I$ and the $\tilde{u}$'s are
given by
\begin{equation}\label{1pX2}
w \; y^I = - \; (1 + w) \; \tilde{u}^0 - \; (1 - w) \; \left(
\sum_{i \in \parallel_I} u^i - \frac{U^c}{m} \right) 
\end{equation}
\begin{eqnarray}
\tilde{u}^0_{\tilde{\sigma} \tilde{\sigma}} 
+ \alpha \tilde{u}^0_{\tilde{\sigma}} 
+ \tilde{s}^0 \; (\tilde{u}^0_{\tilde{\sigma}} 
- \tilde{u}_{\tilde{\sigma}}) & = &
\sum_I \tilde{c}^0 \; \frac{R}{4} \; (y^I + 2 \tilde{u}) 
\label{10X2} \\
\tilde{u}^i_{\tilde{\sigma} \tilde{\sigma}} 
+ \alpha \tilde{u}^i_{\tilde{\sigma}} & = & 
\sum_I \tilde{c}^{i I} \;
\frac{R}{4} \; (y^I + 2 \tilde{u}) \label{1iX2}
\end{eqnarray}
\begin{equation}\label{1quadX2}
\tilde{u}^0_{\tilde{\sigma}} \; = \; 
\sum_I \frac{w}{m} \; \frac{R}{4} \; (y^I + 2 \tilde{u}) 
+ (m - 1) \; e^{2 \tilde{\lambda}_0} \; \tilde{u} 
\end{equation}
\begin{equation}\label{1sigmaX2}
\tilde{u}^0_{\tilde{\sigma}} - \tilde{u}_{\tilde{\sigma}} 
\; = \; \sum_I \left( \frac{w - 1}{m} \right) \; \frac{R}{4} \;
(y^I + 2 \tilde{u}) + (m - 1) \; e^{2 \tilde{\lambda}_0} \; 
(2 \tilde{u}) \; \; .
\end{equation}

Note that $y^I$ depends on $u^i$, and also on $I$ through the
sum $\sum_{i \in \parallel_I} u^i \;$. However, with $w^I = w$
for all $I$, the equations for $\tilde{u}^0$ and $\tilde{u}$
depend not on the individual $y^I$ but only on their sum $\sum_I
y^I \;$. Now, for the ${\cal N} = 4$, $2 2' 5 5' \; : \; (12,
34, 13567, 24567)$ configuration, we have $m = 2$ and
\[ 
\sum_I \left( \sum_{i \in \parallel_I} u^i - \frac{U^c}{m}
\right) \; = \; 2 \; U^c - {\cal N} \; \frac{U^c}{m} \; = \; 0
\; \; . 
\]
Therefore, for the $2 2' 5 5'$ configuration, the sum
$\sum_Iy^I$ is independent of $u^i \;$. Then 
\[
\frac{R}{4} \; \sum_I y^I = - \; \frac{1 + w}{w} \; R \;
\tilde{u}^0 
\]
and the equations for $\tilde{u}^0$ and $\tilde{u}$ become the
same as those in Example (1). In particular, it follows that the
perturbations exhibit damped oscillations and, hence, there is
instability. Therefore, sufficiently massive stars made up of $2
2' 5 5'$ intersecting branes will collapse.

Equation for $\tilde{u}^i$ depend on $I$ through the term
$\sum_I \tilde{c}^{i I} \; (y^I + 2 \tilde{u}) \;$, and hence
through the sums $\sum_I \tilde{c}^{i I}$ and $\sum_I
\tilde{c}^{i I} y^I \;$. Such sums for the $2 2' 5 5'$
configuration are given by the expressions given in equation
(\ref{ciXi}), from which it can be seen that
\[
\sum_I \tilde{c}^{i I} \; = \; 0 
\; \; \; \Longrightarrow \; \; \;
\sum_I \tilde{c}^{i I} \; y^I \; = \; \sum_I \tilde{c}^{i I} \;
\left( \sum_{j \in \parallel_I} u^j \right) \; \; .
\]


For $I : (2, 2', 5, 5')$, the sums $\sum_{j \in \parallel_I}
u^j$ are given by $(u^1 + u^2)$, $\; (u^3 + u^4)$, $\; (u^1 +
u^3 + u^5 + u^6 + u^7)$, and $(u^2 + u^4 + u^5 + u^6 + u^7) \;$.
We then get
\[
\tilde{u}^i_{\tilde{\sigma} \tilde{\sigma}} 
+ \alpha \tilde{u}^i_{\tilde{\sigma}} \; = \; \frac{R}{4} \;
\sum_I \tilde{c}^{i I} \; y^I
\]
where 
\begin{eqnarray} 
\sum_I \tilde{c}^{1 I} \; y^I & = & 2 \; 
\tilde{c}^\parallel \; (u^1 - u^4) \nonumber \\
\sum_I \tilde{c}^{2 I} \; y^I & = & 2 \; 
\tilde{c}^\parallel \; (u^2 - u^3) \nonumber \\
\sum_I \tilde{c}^{3 I} \; y^I & = & 2 \; 
\tilde{c}^\parallel \; (u^3 - u^2) \nonumber \\
\sum_I \tilde{c}^{4 I} \; y^I & = & 2 \; 
\tilde{c}^\parallel \; (u^4 - u^1) \nonumber \\
\sum_I \tilde{c}^{5, 6, 7 \;  \; I} \; y^I & = & 2 \; 
\tilde{c}^\parallel \; (u^5 + u^6 + u^7) \; \; . \label{rhsui}
\end{eqnarray}
It then follows easily that 
\[
(*_1)_{\tilde{\sigma} \tilde{\sigma}} + \alpha
(*_1)_{\tilde{\sigma}} - R \; \tilde{c}^\parallel \; (*_1)
\; = \; 0 
\]
where $(*_1) = (\tilde{u}^1 - \tilde{u}^4)$, $\; (\tilde{u}^2 -
\tilde{u}^3)$, and $\; (u^5 + u^6 + u^7) \;$; and
\[
(*_2)_{\tilde{\sigma} \tilde{\sigma}} + \alpha
(*_2)_{\tilde{\sigma}} \; = \; 0 
\]
where $(*_2) = (\tilde{u}^1 + \tilde{u}^4)$, $\; (\tilde{u}^2 +
\tilde{u}^3)$, $\; (\tilde{u}^5 + \tilde{u}^6 - 2 \tilde{u}^7)$,
and $\; (\tilde{u}^5 - 2 \tilde{u}^6 + \tilde{u}^7) \;$.


\vspace{4ex}

\centerline{\bf $\; {\mathbf {\cal N} = 3} \;$
case -- $\; {\mathbf 2 2' 2''} \;$ configuration}

\vspace{2ex} 

We now consider the $2 2' 2'' \; : \; (12, 34, 56)$
configuration of the M theory where there are three stacks of
two branes whose spatial worldvolume directions are indicated.
For this configuration, we have ${\cal N} = 3$, $\; n_c = 6$,
and $m = 3 \;$. Further analysis proceeds in the same way as for
the ${\cal N} = 4$ case. The steps differ in details, but the
final results for the ${\cal N} = 3$ case of $2 2' 2''$
intersecting brane configuration are almost identical to those
of the ${\cal N} = 4$ case, but now with $m = 3 \;$.

Thus: $s^i = 0$; the same zeroth order results as in Example (1)
are obtained; and, the equations for $\tilde{u}^0$ and
$\tilde{u}$ become the same as those in Example (1). In
particular, it follows that the perturbations exhibit damped
oscillations and, hence, there is instability. Therefore,
sufficiently massive stars made up of $2 2' 2''$ intersecting
branes will collapse.


\vspace{4ex}

\centerline{\bf 6. Significance of (in)stabilities 
-- Mathur's fuzzball proposal} 

\vspace{2ex}

We now discuss the significance of the (in)stability results for
the stars in M theory, within the context of Mathur's fuzzball
proposal \cite{fuzz}. We draw a corollary of this proposal,
which is perhaps obvious to many readers; discuss its
implication for the equations considered here; and make a few
comments.
 
We now have that stars in M theory made up of $M2$ or $M5$
branes, or the ${\cal N} = 4$ $\; 2 2' 5 5'$ intersecting
branes, or the ${\cal N} = 3$ $\; 2 2' 2''$ intersecting branes
all have instabilities. Therefore, if sufficiently massive, such
stars in M theory will collapse presumably to form black brane
configurations. The results of \cite{pc07} that stable,
arbitrarily massive, equilibrium configurations of stars may
exist if $m \ge 9 \;$ are not applicable to the stars in M
theory made up of branes and intersecting branes because $D =
11$ and $n_c + m = 9$ in M theory, and $n_c > 1$ for branes and
intersecting branes.

From many points of view, such a result is only to be
expected. However, consider Mathur's fuzz ball proposal
\cite{fuzz}. According to this proposal, in essence, there is no
horizon. \footnote{ Several horizonless pictures have been
advocated from several different points of view over time. For a
sample of such works, see \cite{kg} -- \cite{page}.} At
distances greater than of the order of Schwarzschild radius, the
spacetime geometry is indistinguishable from a black hole /
black brane geometry but, at shorter distances, the spacetime is
filled with `fuzz'.

\vspace{1ex}

It then follows as a corollary of the fuzz ball proposal that

\begin{itemize}

\item

{\em The equilibrium configurations of stars of arbitrarily high
mass must exist which must be stable and not collapse}. 

\end{itemize}

\noindent 
This must be true irrespective of whether the stars are neutral
or charged, rotating or non rotating. In the following, we
proceed with the assumption that the fuzz ball proposal and,
consequently, its corollary above are valid.

\vspace{1ex}

We now make a remark regarding fuzz ball configurations. A
generic fuzz ball state is highly quantum and has sizeable
fluctuations of metric and other fields. Its connection to
classical geometry is very much more involved than the standard
quantum to classical correspondence. Here, we envision the
fuzzball configurations as describing intersecting branes with
all the high energy, highly interacting excitations included.
Our analysis requires only the energy momentum tensor and the
equations of state describing such configurations. They may be
derived in certain approximations as in \cite{cm} or by using
symmetries as in \cite{k07}. Or they may be postulated based on
underlying microphysics if known, or based simply on educated
guesswork.

The fuzz ball situation here may be likened to studying stars
made up of photons, or nuclear matter, or Bose -- Einstein
condensates. These constituents are highly quantum.
Nevertheless, to study the properties of such stars, the
equations of state for the constituents are sufficient. For
photons, one has $ (pressure) = (density) / 3 \;$; for nuclear
matter, one postulates different equations of states; and for
Bose -- Einstein condensates, they may be derived from
statistical mechanics.

It is important to find the equilibrium horizonless
configurations implied by the fuzz ball proposal, and understand
the reasons for their stability. One may therefore seek
stationary configurations since equilibrium ones are in general
of this type. If one further assumes no rotation then one may
restrict to static and spherically symmetric configuration. Then
one may proceed, for example, by following the methods of
\cite{cpaper, p, pc02, pc07} as done in this paper : First
obtain OV type equations corresponding to the equilibrium
configurations of the `fuzz'; derive or postulate appropriate
equations of state for the `fuzz', which may atleast capture its
stabilising aspects; then obtain the asymptotic solutions to
these equations; study the behaviour of perturbations in the
mass of the star around the asymptotic solutions; and, determine
whether an instability is present or not depending on whether
perturbations are oscillatory or not.

The OV type equations, and the equations of state, given in this
paper for stars in M theory predict instabilities. Therefore,
the assumed validity of the corollary above implies that
although these equations may describe well the stars in M
theory, they are not sufficiently general to describe the `fuzz'
which must be able to prevent the collapse. We suspect that, at
the least, the isometries along the compact space needs to be
removed or modified. Note that the absence of such isometries
invalidates our use of U duality symmetries in \cite{k07} to
obtain their effects on the equations of state. One then has to
postulate, or derive by other means, equations of state for the
`fuzz' which must be able to prevent the collapse. This,
however, is a tall order and it is presently not clear to us how
to proceed.


\vspace{4ex}

\centerline{\bf 7. Conclusions} 

\vspace{2ex}

We summarise briefly the results presented in this paper.  Our
aim is to study the stars in M theory and determine whether they
are stable or not. We start with appropriate higher dimensional
spacetime, applicable to stars in M theory. We obtain the analog
of OV equations. The equations of state are taken to be
linear. We then follow the standard methods to determine the
presence or absence of instabilities : we obtain analytically
the asymptotic solutions and study perturbations around it.

A general answer is algebraically quite involved. Hence, we
consider a few examples. In the process, we obtain a
generalisation of the results of Chavanis given in \cite{pc07}.
We also obtained the results for stars in M theory made up of
$M2$ and $M5$ branes; and the ${\cal N} = 4$ $\; 2 2' 5 5'$
intersecting branes. The results for ${\cal N} = 3$ $\; 2 2'
2''$ intersecting branes are similar to the ${\cal N} = 4$
case. All these stars in M theory have instabilities. Therefore,
if sufficiently massive, they will collapse to form black brane
configurations. We then discussed the significance of these
(in)stabilities within the context of Mathur's fuzz ball
proposal.

We now conclude by mentioning a few topics for further studies.
With the OV type equations given here, one can study stars in M
theory, and also stars in more general Kaluza -- Klein
spacetimes. The equations of state for the star's constituents
in the later spacetimes may need to be postulated. However, one
can use this freedom and explore the consequences of various
types of equations of state.

One can also study the collapse of such stars, and obtain the
analog of Oppenheimer -- Snyder collapse scenario. This is
likely to be an interesting area of research because of its
similarities to the time dependent cosmological studies. It will
be interesting to explore the similarities between big bang /
big crunch scenarios and the collapse of a star down the
singularity to form a black hole. And, in particular, explore
whether a bounce in the scale factor during the early universe
evolution implies a bounce for the collapsing star and, if so,
whether it leads to stability.

At a technical level, we have not analysed all the intersecting
brane configurations in full generality. For example, in all the
cases here, we end up with $w^I = w \;$. This needs
generalisation. Also, one can consider more general equations of
state and explore the possible consequences. This will require
numerical analysis and is likely to be challenging. However,
this may yet be the most practical way to model the `fuzz' and
to find the stable equilibrium configurations.




\vspace{4ex}

\centerline{\bf Appendix A : Equations of motion in a more
compact form}

\vspace{2ex}

In this Appendix, we write the equations of motion (\ref{Pir})
-- (\ref{alpharr}) in a more compact form. Note that equation
(\ref{alpharr}) suggests a change of variable from $r$ to $\tau$
given by
\begin{equation}\label{artau}
e^\lambda \; d r = e^\Lambda \; d \tau \; \; .
\end{equation}
The line element for the $(m + 1)$ dimensional transverse space
given in equation (\ref{ds}), written in terms of $\tau$,
becomes
\begin{equation}\label{adsm+1} 
e^{2 \lambda} d r^2 + e^{2 \sigma} d \Omega_m^2 \; = \; 
e^{2 \sigma} \; \left( e^{2 \chi} \; d \tau^2 + d \Omega_m^2
\right) 
\end{equation}
where the field $\chi$ is defined by 
\begin{equation}\label{achidefn}
\chi = \Lambda - \sigma = 
\lambda^0 + \sum_i \lambda^i + (m - 1) \sigma \; \; . 
\end{equation}
It follows from the above definitions that
\begin{equation}\label{artauf} 
r_\tau = \frac{d r}{d \tau} = e^{\Lambda - \lambda}
= r \; \sqrt{f} \; e^\chi 
\end{equation}
and that, for any function $X(r(\tau))$, 
\begin{equation}\label{aXtaur}
X_\tau = e^{\Lambda - \lambda} \; X_r
\; \; , \; \; \; 
X_{\tau \tau} = e^{2 (\Lambda - \lambda)} \; \left( X_{r r} 
+ (\Lambda_r - \lambda_r) \; X_r \right)
\end{equation}
where the subscripts $\tau$ denote $\tau-$derivatives.
Equations (\ref{Pir}) -- (\ref{alpharr}), expressed in terms of
$\tau$, become
\begin{eqnarray}
(\Pi_I)_\tau & = & - \Pi_I \; \Lambda_\tau 
+ \sum_\alpha p_{\alpha I} \lambda^\alpha_\tau \label{aPitau} \\
\Lambda^2_\tau - \sum_\alpha (\lambda^\alpha_\tau)^2 & = &
2 \sum_I \Pi_I \; e^{2 \Lambda} 
+ m (m - 1) \; e^{2 \chi} \label{aLambdatau2} \\
\lambda^\alpha_{\tau \tau} 
& = & \sum_I \left(- p_{\alpha I} + \frac{T_I}{D - 2} \right) \;
e^{2 \Lambda} + \delta^{\alpha a} \; (m - 1) \; e^{2 \chi}
\; \; . \label{aalphatautau}
\end{eqnarray}
Equation for the field $\chi$ defined in equation
(\ref{achidefn}) is then given by
\begin{equation}\label{aL-sigmatautau}
\chi_{\tau \tau} = \sum_I \left( \Pi_I + p_{\Omega I} \right) \;
e^{2 \Lambda} + (m - 1)^2 \; e^{2 \chi} \; \; .
\end{equation}
Defining $G_{\alpha \beta}$, $\; G^{\alpha \beta}$, and
$h_{\alpha I}$ by
\begin{eqnarray}
G_{\alpha \beta} & = & 1 - \delta_{\alpha \beta} 
\; \; \; , \; \; \; \;
G^{\alpha \beta} \; = \; \frac{1}{D - 2} - \delta^{\alpha \beta}
\; \; \; , \nonumber \\
& & \nonumber \\
h_{\alpha I} & = & \sum_\beta G_{\alpha \beta} \; 
\left(- p_{\beta I} + \frac{T_I}{D - 2} \right) \; = \;
\Pi_I + p_{\alpha I} \; \; ,  \label{aghalphaI}
\end{eqnarray}
the above equations may be written compactly as
\begin{eqnarray}
(\Pi_I)_\tau & = & - 2 \Pi_I \; \Lambda_\tau 
+ \sum_\alpha h_{\alpha I} \lambda^\alpha_\tau \label{a2Pitau} \\
\sum_{\alpha \beta} G_{\alpha \beta} \; \lambda^\alpha_\tau \;
\lambda^\beta_\tau & = & 2 \sum_I \Pi_I \; e^{2 \Lambda} 
+ m (m - 1) \; e^{2 \chi} \label{a2Lambdatau2} \\
\lambda^\alpha_{\tau \tau} & = & 
\sum_{\beta I} G^{\alpha \beta} \; h_{\beta I} \; 
e^{2 \Lambda} + \delta^{\alpha a} \; (m - 1) \; 
e^{2 \chi} \; \; . \label{a2alphatautau} \\
\chi_{\tau \tau} & = & \sum_I h_{\Omega I} \; e^{2 \Lambda} 
+ (m - 1)^2 \; e^{2 \chi} \; \; .  \label{a2L-sigmatautau}
\end{eqnarray}


\vspace{4ex}

\centerline{\bf Appendix B : Linear equations of state} 

\vspace{2ex}

When the equations of state are linear, the fields
$(\lambda^\alpha, \; \rho_I)$ can be expressed in terms ${\cal
N} + 1$ independent fields, denoted as $(l^I, \; l^*) \;$.
Consider the linear equations of state given by equations
(\ref{eoswi}), namely
\begin{equation}\label{aeoswi}
p_{\alpha I} = w^I_\alpha \; \rho_I \; \; , \; \; \; 
\Pi_I = w^I_\Pi \; \rho_I
\end{equation}
where $(w^I_\alpha, \; w^I_\Pi)$ are constants, $w^I_0 = - 1$
since $p_{0 I} = - \rho_I$, and $w^I_a = w^I_\Omega = w^I$ since
$p_{a I} = p_{\Omega I} = p_I$ for all $a \;$. Then 
\begin{equation}\label{aeos}
h_{\alpha I} = \Pi_I + p_{\alpha I} = z^I_\alpha \; \Pi_I
\end{equation}
where $z^I_\alpha$ are constants and $z^I_a = z^I_\Omega$ for
all $a \;$. The sets of constants $z^I_\alpha$ and $(w^I_\alpha,
\; w^I_\Pi)$ where $w^I_0 = - 1$ are related to each other as
follows:
\[
z^I_\alpha = 1 + \frac{w^I_\alpha}{w^I_\Pi}
\; \; \; ; \; \; \; \;   \; \; \; 
w^I_\alpha = \frac{z^I_\alpha - 1}{1 - z^I_0} 
\; \; , \; \; \;
w^I_\Pi = \frac{1}{1 - z^I_0} \; \; .
\]

Equation (\ref{a2Pitau}) can now be solved using equation
(\ref{aeos}) to give
\begin{equation}\label{aPiIhI} 
\Pi_I = \Pi_{I 0} \; e^{- 2 \Lambda + l^I} 
\; \; \; \longrightarrow \; \; \; 
h_{\alpha I} \; e^{2 \Lambda} = z^I_\alpha \; 
\left( \Pi_{I 0} \; e^{l^I} \right)
\end{equation}
where $\Pi_{I 0}$ are constants and $l^I$ are defined by
\begin{equation}\label{alIdefn} 
l^I = \sum_\alpha z^I_\alpha \; \lambda^\alpha \; \; . 
\end{equation}
An expression for $\rho_I$ follows easily since $\Pi_I = w^I_\Pi
\; \rho_I \;$. Using the above expressions now, equations for
$\lambda^\alpha$, $\; l^I$, and $\chi$ become
\begin{eqnarray}
\lambda^\alpha_{\tau \tau} & = & \sum_I z^{\alpha I} \; 
\left( \Pi_{I 0} \; e^{l^I} \right)
+ \delta^{\alpha a} \; (m - 1) \; e^{2 \chi} \label{alalphatt} \\
l^I_{\tau \tau} & = & \sum_J {\cal G}^{I J} \; 
\left( \Pi_{J 0} \; e^{l^J} \right) 
+ z^I_\Omega \; m (m - 1) \; e^{2 \chi} \label{alItt} \\
\chi_{\tau \tau} & = & \sum_I z^I_\Omega \; 
\left( \Pi_{I 0} \; e^{l^I} \right)
+ (m - 1)^2 \; e^{2 \chi}  \label{achitt}
\end{eqnarray}
where 
\begin{equation}\label{aGIJ}
z^{\alpha I} =  \sum_\beta G^{\alpha \beta} \; z^I_\beta
\; \;, \; \; \; 
{\cal G}^{I J} = \sum_{\alpha \beta} 
G^{\alpha \beta} z^I_\alpha \; z^J_\beta \; \; . 
\end{equation}

The above set of equations may be written even more compactly,
and $\lambda^\alpha$ may be expressed in terms of $(l^I, \chi)$,
by defining the following:
\[
\hat{I} = (I, *) \; \; , \; \; \; 
l^{\hat{I}} = (l^I, l^*) \; \; , \; \; \; 
z_\alpha^{\hat{I}} = (z_\alpha^I, z_\alpha^*)
\; \; , \; \; \; 
\Pi_{\hat{I} 0} = (\Pi_{I 0}, \Pi_{* 0})
\]
where $z_a^{\hat{I}} = z_\Omega^{\hat{I}} = (z_\Omega^I,
z_\Omega^*) $ for all $a \;$, and
\[
l^* = 2 \chi = \sum_\alpha z_\alpha^* \; \lambda^\alpha 
\; \; , \; \; \; 
z_\alpha^* = 2 \left( 1 - \frac{\delta_{\alpha a}}{m} \right)
\; \; , \; \; \; 
\Pi_{* 0} = \frac{m (m - 1)}{2} \; \; . 
\]
Equations (\ref{alalphatt}), (\ref{alItt}) and (\ref{achitt})
may then be written as
\begin{eqnarray}
\lambda^\alpha_{\tau \tau} & = & \sum_{\hat{J}} 
z^{\alpha \hat{J}} \; \left( \Pi_{\hat{J} 0} \; e^{l^{\hat{J}}}
\right) \label{alambdahat} \\
l^{\hat{I}}_{\tau \tau} & = & \sum_{\hat{J}} 
\hat{{\cal G}}^{\hat{I} \hat{J}} \; 
\left( \Pi_{\hat{J} 0} \; e^{l^{\hat{J}}} \right) 
\label{alhattt} 
\end{eqnarray}
where 
\begin{equation}\label{ahatGIJ}
z^{\alpha \hat{I}} = \sum_\beta G^{\alpha \beta} \;
z^{\hat{I}}_\beta \; \; , \; \; \;
\hat{{\cal G}}^{\hat{I} \hat{J}} = \sum_{\alpha \beta} 
G^{\alpha \beta} z^{\hat{I}}_\alpha \; z^{\hat{J}}_\beta \; \; .
\end{equation}
It follows that $z^{\alpha *} = \frac{2}{m} \; \delta^{\alpha
a}$, and that $\hat{{\cal G}}^{\hat{I} \hat{J}}$ are given by
\[
\hat{{\cal G}}^{I J} = {\cal G}^{I J} 
\; \; , \; \; \; 
\hat{{\cal G}}^{\hat{I} *} =
\hat{{\cal G}}^{* \hat{I}} = n^{\hat{I}} 
= (n^I, n^*) = 2 \; (z_\Omega^I, z_\Omega^*) \; \; . 
\]
Thus 
\[
\hat{{\cal G}}^{* I} = \hat{{\cal G}}^{I *} = 2 \; z_\Omega^I
\; \; , \; \; \; 
\hat{{\cal G}}^{* *} = 2 \; z_\Omega^* = \frac{4 (m - 1)}{m} 
\; \; .
\]
When $l^{\hat{I}}$ are all independent, equation (\ref{alhattt})
can be inverted to give
\begin{equation}\label{apihat0}
\Pi_{\hat{I} 0} \; e^{l^{\hat{I}}} = \sum_{\hat{J}} 
\hat{{\cal G}}_{\hat{I} \hat{J}} \; l^{\hat{J}}_{\tau \tau}
\end{equation}
where $\hat{{\cal G}}_{\hat{I} \hat{J}}$ are the elements of the
inverse of the matrix formed by $\hat{{\cal G}}^{\hat{I}
\hat{J}} \;$, and are given by
\[
\hat{{\cal G}}_{I J} = {\cal G}_{I J} + \frac{n_I \; n_J}{N}
\; \; , \; \; \; 
\hat{{\cal G}}_{I *} = \hat{{\cal G}}_{* I} = - \; \frac{n_I}{N}
\; \; , \; \; \; 
\hat{{\cal G}}_{* *} = \frac{1}{N} \; \; . 
\]
Here ${\cal G}_{I J}$ are the elements of the inverse of the
matrix formed by ${\cal G}^{I J}$ and
\[
n_I = \sum_J {\cal G}_{I J} \; n^J
\; \; , \; \; \; 
N = n^* - \sum_J n_J \; n^J \; \; . 
\]
From equations (\ref{alambdahat}) and (\ref{apihat0}), it now
follows that
\begin{eqnarray}
\lambda^\alpha & = &  \sum_{\hat{I} \hat{J}} z^{\alpha \hat{I}} \;
\hat{{\cal G}}_{\hat{I} \hat{J}} \; l^{\hat{J}} 
+ L^\alpha \; (\tau - \tau_0) + c^\alpha \nonumber \\
& & \nonumber \\
& = &  \sum_J z^\alpha_J \; l^J + z^\alpha_* \; l^*
+ L^\alpha \; (\tau - \tau_0) + c^\alpha \label{a2lambda}
\end{eqnarray}
where $(L^a, c^a) = (L^\Omega, c^\Omega)$ for all $a$, 
\[
z^\alpha_{\hat{J}} = (z^\alpha_J, z^\alpha_*) = \sum_{\hat{I}}
z^{\alpha \hat{I}} \; \hat{{\cal G}}_{\hat{I} \hat{J}} 
= \sum_{\beta \hat{I}} G^{\alpha \beta} \; z^{\hat{I}}_\beta \;
\hat{{\cal G}}_{\hat{I} \hat{J}} \; \; ,
\]
and $(L^\alpha, c^\alpha)$ are constants obeying the following
constraints:
\begin{equation}\label{aLconstrnt}
l^{\hat{I}} = \sum_\alpha z_\alpha^{\hat{I}} \; \lambda^\alpha
\; \; \; \Longrightarrow \; \; \;
\sum_\alpha z_\alpha^{\hat{I}} \; L^\alpha = \sum_\alpha
z_\alpha^{\hat{I}} \; c^\alpha = 0 \; \; . 
\end{equation}
Substituting the expression for $\lambda^\alpha$ from equation
(\ref{a2lambda}) into equation (\ref{a2Lambdatau2}) and after
some algebra, it follows that
\begin{equation}\label{alquad}
\sum_{\hat{I} \hat{J}} \hat{{\cal G}}_{\hat{I} \hat{J}} \;
l^{\hat{I}}_\tau \; l^{\hat{J}}_\tau
= 2 \; \sum_{\hat{I}} \Pi_{\hat{I} 0} \; e^{l^{\hat{I}}} 
- \sum_{\alpha \beta} G_{\alpha \beta} L^\alpha L^\beta \; \; .
\end{equation}
Note that 
\[
\sum_\alpha z_\alpha^* \; L^\alpha = 0 
\; \; \; \Longrightarrow \; \;\; \; 
\sum_\alpha L^\alpha = L^\Omega  
\]
and, hence, that 
\begin{eqnarray*}
- \sum_{\alpha \beta} G_{\alpha \beta} L^\alpha L^\beta & = &
\sum_\alpha (L^\alpha)^2 - (\sum_\alpha L^\alpha)^2 \\ 
& = & (L^0)^2 + \sum_i (L^i)^2 + (m - 1) (L^\Omega)^2
\; \ge \; 0 \; \; .
\end{eqnarray*}
One may now solve equations (\ref{alhattt}) and (\ref{alquad})
and obtain $l^{\hat{I}} (\tau) \;$. Equation (\ref{a2lambda})
then gives $\lambda^\alpha (\tau) \;$. One now has to make a
choice of $r$ using the diffeomorphic freedom in defining it. In
this paper, $r$ is chosen to be given by $r = e^{\tilde{\sigma}}
= e^{\sigma + \frac{\Lambda^c}{m}} \;$. Equation (\ref{artauf})
then gives $\sqrt{f} = \tilde{\sigma}_\tau e^{- \chi}$,
thereby completing the solutions.

It is easy to see that, when written in terms of the variable
$r$, $\; (\tau - \tau_0)$ in equation (\ref{a2lambda}) for
$\lambda^\alpha$ is to be replaced by a function $F(r)$ which
obeys
\[
F_{r r} + (\Lambda_r - \lambda_r) \; F_r = 0 
\; \; \; \Longleftarrow \; \; \; 
F_{\tau \tau} = 0 \; \; . 
\]
It is also easy to see that $\tilde{\lambda}^\alpha =
\lambda^\alpha + \frac{\Lambda^c}{m}$ are given by
\begin{eqnarray}
\tilde{\lambda}^\alpha & = &  
\sum_{\hat{I} \hat{J}} \tilde{z}^{\alpha \hat{I}} \;
\hat{{\cal G}}_{\hat{I} \hat{J}} \; l^{\hat{J}} 
+ \tilde{L}^\alpha \; (\tau - \tau_0) + \tilde{c}^\alpha \nonumber \\
& & \nonumber \\
& = &  \sum_J \tilde{z}^\alpha_J \; l^J + \tilde{z}^\alpha_* \; l^*
+ \tilde{L}^\alpha \; (\tau - \tau_0) + \tilde{c}^\alpha 
\label{atld2lambda}
\end{eqnarray}
where $ \tilde{z}^\alpha_{\hat{J}} = (\tilde{z}^\alpha_J,
\tilde{z}^\alpha_*) = \sum_{\hat{I}} \tilde{z}^{\alpha \hat{I}}
\; \hat{{\cal G}}_{\hat{I} \hat{J}} \;$ and
\[
\tilde{X}^\alpha = X^\alpha + \frac{\sum_i X^i}{m} 
\; \; , \; \; \; 
X^\alpha = (z^{\alpha \hat{I}}, \; L^\alpha, \; c^\alpha) 
\; \; .
\]
For example, using $z^{\alpha \hat{I}} = \sum_\beta G^{\alpha
\beta} \; z^{\hat{I}}_\beta \;$ and $\sum_\alpha L^\alpha
=L^\Omega \;$, it follows that
\[
\tilde{z}^{\alpha \hat{I}} = - z^{\hat{I}}_\alpha 
+ \frac{z^{\hat{I}}_0 + m z^{\hat{I}}_\Omega}{m}
\; \; \; , \; \; \; \; 
\tilde{L}^\alpha = L^\alpha - \frac{L^0 + (m - 1) L^\Omega}{m}
\; \; .
\]


\vspace{4ex}

\centerline{\bf Appendix C : M theory branes}

\vspace{2ex}

For intersecting branes in M theory, the U duality relations
further lead to an elegant structure. Firstly, it can be argued
\cite{k07} that the equations of state may be written in terms
of one single function only. For example, they may be written as
\begin{equation}\label{aueos}
h_{\alpha I} = \Pi_I + p_{\alpha I} = 
z_\alpha \; {\cal F} (\{*\}_I)
\; \; , \; \; \;
\alpha = \{ 0, \parallel, \perp, \Omega \} 
\end{equation}
where $\{*\}_I$ denote brane quantities such as the number and
the nett charge of the branes in $I^{th}$ stack, the constant
coefficients $z_\alpha$ and the functional dependence of ${\cal
F}$ on the brane quantities $\{*\}$ are same for all $I \;$, and
\begin{equation}\label{az}
z_\Omega = z_\perp \; \; \; , \; \; \; \; 
z_\parallel = z_0 + z_\perp 
\end{equation}
as follows from $p_{\Omega I} = p_{\perp I}$ and from equation
(\ref{Ups}). Note that equations (\ref{aueos}) may still lead to
equations of state of the type considered in equation
(\ref{aeos}) if ${\cal F}$ depends on $\Pi$ only and if the
function ${\cal F}(\Pi)$ is such that, for example, ${\cal
F}(\Pi) = u^{(lo)} \Pi \; \;$ or $\; u^{(hi)} \Pi \; \;$ for low
or high magnitudes of $\Pi \;$. Equation (\ref{aeos}) then
follows, with $z^I_\alpha = z_\alpha u^I$ where $u^I = u^{(lo)}$
or $u^{(hi)}$ according to whether the magnitude of $\Pi_I$ is
low or high.

Consider the case where $z^I_\alpha = z_\alpha u^I \;$ and
$z_\alpha$ obey equations (\ref{az}). It then follows from
equations (\ref{aGIJ}) and from BPS intersection rules that
\begin{equation}\label{agenz^alphaI}
z^{\alpha I} = \left( \frac{(n_I + 1)}{9} \; z_0 + z_\perp 
- z_\alpha \right) \; u^I 
\end{equation}
\begin{equation}\label{agenGIJM}
{\cal G}^{I J} = 2 z_0 
\left( z_\perp - z_0 \; \delta^{I J} \right) \; u^I u^J \; \; .
\end{equation}
The form of $z^{\alpha I}$ and the relations among them are
consequences of U duality symmetries of M theory. The relations
among $z^{\alpha I}$ will be reflected in the solutions for
$\lambda^\alpha$ when written in terms of $l^I$, see equation
(\ref{a2lambda}). Note the elegant structure of ${\cal G}^{I J}$
and its independence on $n_I \;$. These features of ${\cal G}^{I
J}$ are consequences of both the U duality symmetries and the
BPS intersection rules. Similar ${\cal G}^{I J}$ was also
present in our time dependent early universe studies.

Note that if ${\cal G}^{I J}$ are of the form 
\begin{equation}\label{agenGIJ}
{\cal G}^{I J} = a \left( b + \delta^{I J} \right) \; u^I u^J
\end{equation}
then ${\cal G}_{I J}$ are given by 
\begin{equation}\label{ainvgenGIJ}
{\cal G}_{I J} = \frac{\beta + \delta_{I J}}{a \; u^I u^J} 
\; \; \; , \; \;\; \; \; 
\beta = - \; \frac{b}{{\cal N} b + 1} \; \; . 
\end{equation}
Thus, for the BPS intersecting branes in M theory, it follows
from equation (\ref{agenGIJM}) that
\begin{equation}\label{ageninvGIJM}
{\cal G}_{I J} = \frac{1}{2 z_0^2 \; u^I u^J} \; 
\left( \frac{z_\perp}{{\cal N} z_\perp - z_0} - \delta_{I J}
\right) \; \; .
\end{equation}



\begin{thebibliography}{999}
\bibitem{fn1}
Rigorously, what is implied by the relation $2 w^I = (1 + w^I)
\tilde{s}^0$ is the following: When $\tilde{c}^{i I} = 0$,
equation (\ref{phii}) implies that $\rho_I \propto e^{- \frac{1
+ w^I} {w^I} \; \tilde{\lambda}^0} \;$. If $w^I$s are different,
then $\rho_I$s will also be different and, in the limit $r \to
\infty \;$, some of them will be dominating the others. In such
a case, in an asymptotic analysis, the subdominant $\rho_I$s
must be ignored and only the dominant ones kept. This implies
that the value of ${\cal N}$ is effectively lowered; and that
all the $\rho_I$s will participate in the dynamics if and only
if $w^I$ are all equal.
\end{thebibliography}
\end{document}